\documentclass[journal]{IEEEtran}
\usepackage[final]{graphicx}
\usepackage{subfigure}
\usepackage{float}
\usepackage{amsmath}
\usepackage{cases}
\usepackage{color}
\usepackage{colortbl}
\usepackage{algorithm}
\usepackage{algpseudocode}
\usepackage{amsmath}
\usepackage{amssymb}
\usepackage{booktabs}
\usepackage{setspace}
\usepackage{array}
\usepackage{soul}
\usepackage{bbding}

\newtheorem{theorem}{\noindent \bf Theorem}
\newenvironment{proof}{{ \it Proof:}}{\hfill $\blacksquare$\par}

\makeatletter
\renewcommand{\maketag@@@}[1]{\hbox{\m@th\normalsize\normalfont#1}}%
\makeatother

\usepackage{stfloats}
\usepackage{cite}
\usepackage{makecell}
\usepackage{multirow}
\usepackage{hyperref}

\ifCLASSINFOpdf
\else
\fi

\usepackage{caption}
\usepackage{mathtools}

\begin{document}
\title{Integrated Sensing and Communication Signal Processing Based on Compressed Sensing Over Unlicensed Spectrum Bands}
\author{Haotian~Liu,~\IEEEmembership{Student Member,~IEEE,}
	Zhiqing~Wei,~\IEEEmembership{Member,~IEEE,}
	Fengyun~Li,~Yuewei~Lin,~
	Hanyang~Qu,~\IEEEmembership{Member,~IEEE,}
	Huici~Wu,~\IEEEmembership{Member,~IEEE,}	
	Zhiyong~Feng,~\IEEEmembership{Senior~Member,~IEEE}
	\thanks{This work was supported in part by the National Natural Science Foundation of China (NSFC) under Grant U21B2014 and 62271081, and in part by the National Key Research and Development Program of China under Grant 2020YFA0711302. (\textit{Corresponding author: Zhiqing Wei})}
	\thanks{Haotian Liu, Zhiqing Wei, Fengyun Li, Huici Wu and Zhiyong Feng are with Beijing University of Posts and Telecommunications, Beijing, China 
(emails: haotian\_liu@bupt.edu.cn, weizhiqing@bupt.edu.cn, lfy@bupt.edu.cn, dailywu@bupt.edu.cn, fengzy@bupt.edu.cn).

	Hanyang Qu is with China Mobile Zijin Innovation Institute, Nanjing 210000, Jiangsu, China (e-mail: hanyangqu@bupt.cn).
 
	Yuewei Lin is with Qingdao University of Science and Technology, Qingdao 266061, Shandong, China (e-mail: linyuewei@qust.edu.cn).}}

\maketitle

\begin{abstract}
As a promising key technology of 
6th generation (6G) mobile communication system, 
integrated sensing and communication (ISAC) technology aims to make full use of 
spectrum resources to enable the functional integration of communication and sensing. 
The ISAC-enabled mobile communication system regularly operate in non-continuous spectrum bands due to crowded licensed frequency bands. 
However, the conventional sensing algorithms over non-continuous spectrum bands have disadvantages such as reduced peak-to-side lobe ratio (PSLR) and degraded anti-noise performance. 
Facing this challenge, we propose a high-precision ISAC signal processing algorithm based on compressed sensing (CS) in this paper. 
By integrating the resource block group (RBG) configuration information in 5th generation new radio (5G NR) and channel information matrices, we can dynamically and accurately obtain power estimation spectra. 
Moreover, we employ the fast iterative shrinkage-thresholding algorithm (FISTA) to address the reconstruction problem and utilize K-fold cross validation (KCV) to obtain optimal parameters. 
Simulation results show that the proposed algorithm has lower sidelobes or even zero sidelobes compared with conventional sensing algorithms. Meanwhile, compared with the improved 2D FFT algorithm and conventional 2D FFT algorithm, the proposed algorithms in this paper have a maximum improvement of 54.66 \% and 84.36 \% in range estimation accuracy, and 41.54 \% and 97.09 \% in velocity estimation accuracy, respectively.
\end{abstract}


\IEEEpeerreviewmaketitle

\section{Introduction}\label{se1}

With the emerging typical applications such as smart city and intelligent transportation, 
the integrated sensing and communication (ISAC) in 5th generation-advanced (5G-A) and 6th generation (6G) 
mobile communication system aims to integrate sensing services into mobile communication infrastructure~\cite{zhang2022amr,wei2022toward}.
On one hand, compared with separate design of communication and radar sensing, 
the ISAC system realizes the sharing of hardware and spectrum resources of radar sensing and communication, 
thereby saving spectrum resources and improving hardware efficiency \cite{feng2020joint,fan2021radar,du2022tensor,li2019residual}. 
On the other hand, radar and communication systems have high similarities 
in hardware architecture, antenna structure, and spectrum bands, 
which provide feasibility for the integrated design of communication and sensing. 
Therefore, the research on ISAC system has attracted wide attention in academia and industry. 

Due to the scarcity of spectrum resources and the requirement to improve the utilization of spectrum bands, 
cognitive radio (CR) originally proposed by 
Dr. Joseph Mitola has attracted wide attention \cite{mitola1999cognitive}, where unlicensed spectrum bands are utilized. 
At present, unlicensed spectrum bands are also used in mobile communication systems, 
evolving into long term evolution-unlicensed (LTE-U) and 5G new radio-unlicensed (5G NR-U) \cite{yao2018full}. 
Hence, the ISAC technology enabled mobile communication system will work 
in non-continuous spectrum bands~\cite{gu2023cognitive}.
For the communication of unlicensed users, the non-continuous time-frequency domain resources will inevitably bring about a decrease in the communication rate. However, this is beneficial for the total system spectrum resources utilization. In terms of sensing for unlicensed users, the
non-continuous of spectrum bands will lead to the 
degradation of anti-noise performance, resolution, and deterioration of Fourier sidelobes with 
conventional ISAC signal processing algorithms. 
Hence, enhancing the sensing capabilities in non-continuous spectrum bands face two significant challenges: the decline of signal-to-noise ratio (SNR) and deterioration of Fourier sidelobes.

In view of the challenges existing in the ISAC system over non-continuous spectrum bands, the solutions are primarily focused on signal design and signal processing, which are summarized in Table \ref{tab:new add}. 

\textbf{Signal design:} The orthogonal frequency division multiplex (OFDM) pilot signal, 
as a typical non-continuous signal, 
has been studied in the perspectives of pilot design and pilot optimization.
In terms of pilot design, 
Chen \cite{1015730113.nh} designed a time-frequency continuous pilot structure with 
low parameter estimation error and high anti-noise capability.
However, this method has poor anti-noise performance.
Bao \textit{et al.} \cite{bao2020superimposed} 
exploited superimposed pilots to realize target sensing with limited energy and 
high detection probability.
In terms of pilot optimization,
Liu \textit{et al.} \cite{liu2021cramer} optimized the pilot structure by minimizing  
the Cram\'{e}r-Rao bound in the scenario of single target detection.
Tzoreff \textit{et al.} \cite{Tzoreff2017} optimized the pilot matrix by minimizing the 
maximum eigenvalue of the Cram\'{e}r-Rao bound. 
Both of these researches applied the semi-definition relaxation (SDR) to release 
rank-1 constraint. 
However, SDR-based optimization algorithm performs poorly on multi-objective detection problems. 
To this end,
Huang \textit{et al.} \cite{Huang2022} replaced the rank-1 constraint with a tight and 
smooth approximation, 
and applied the majorization-minimization (MM) method to solve the pilot optimization problem.
The above work focuses on designing and optimizing the signal structure to avoid the problems caused by discontinuity. However, in the unlicensed spectrum bands, the spectrum bands occupancy is dynamic and random, making it difficult to avoid discontinuities through signal design.

\begin{table*}
\centering
\caption{Classification and comparison of studies under unlicensed spectrum bands}
\label{tab:new add}
\resizebox{\textwidth}{!}{%
\begin{tabular}{|c|c|c|c|c|}
\hline
\textbf{\begin{tabular}[c]{@{}c@{}}Categories under non-continuous \\ spectrum bands\end{tabular}} & \textbf{Reference} & \textbf{Superiority }  & \textbf{Shortcoming }& \multicolumn{1}{c|}{\textbf{\begin{tabular}[c]{@{}c@{}}Enlightening for solving challenges \\ in unlicensed spectrum bands\end{tabular}}} \\ \hline
\multirow{5}{*}{Signal design}  & \cite{1015730113.nh} & \begin{tabular}[c]{@{}c@{}}Overcoming sidelobes deterioration;\\ sensing range unchanged;\end{tabular} & Poor anti-noise performance  & \XSolidBrush  \\ \cline{2-5} 
& \cite{bao2020superimposed} & Existence of sidelobes deterioration & Excellent anti-noise performance & \XSolidBrush \\ \cline{2-5} 
& \cite{liu2021cramer,Tzoreff2017} & High accuracy in target estimation & \begin{tabular}[c]{@{}c@{}}Slow convergence and \\ performance loss for multi-target detection\end{tabular}   &  \XSolidBrush \\ \cline{2-5} 
& \cite{Huang2022} & Fast convergence  & High complexity & \XSolidBrush  \\ \cline{2-5}  \hline
\multirow{8}{*}{Signal processing}   & \cite{schweizer2017stepped} & Overcome sidelobes deterioration & Amplify the effects of noise & \Checkmark  \\ \cline{2-5} 
& \cite{sturm2013spectrally} &  High sensing resolution & Reduction of maximum unambiguous range  & \XSolidBrush \\ \cline{2-5} 
& \cite{hakobyan2016novel}    &  \begin{tabular}[c]{@{}c@{}}Overcome deterioration of sidelobes\\ for range estimation\end{tabular}  & \begin{tabular}[c]{@{}c@{}}Hard to overcome deterioration of\\ sidelobes for velocity estimation\end{tabular}    & \XSolidBrush \\ \cline{2-5} 
& \cite{knill2018high}  &   \begin{tabular}[c]{@{}c@{}}Overcome deterioration of sidelobes \\ for range and velocity estimation\end{tabular}    & \begin{tabular}[c]{@{}c@{}}Not applicable in dynamic \\ time-frequency domain scenarios\end{tabular} & \Checkmark  \\ \cline{2-5} 
& \cite{wei2022robust}   & \begin{tabular}[c]{@{}c@{}}Suitable for dynamic\\ time-frequency domain resource scenarios\end{tabular}  &   \begin{tabular}[c]{@{}c@{}}Hard to overcome deterioration of\\ sidelobes for range and velocity estimation\end{tabular}   & \Checkmark  \\ \hline
\end{tabular}}
\end{table*}

\textbf{Signal processing:} 
Schweizer \textit{et al.} \cite{schweizer2017stepped} exploited 
a stepped OFDM signal for sensing, 
which is non-continuous in time-frequency domain. 
Then, the range phase error due to discontinuity is compensated by modifying the 
standard discrete Fourier transform (DFT).
However, the method amplifies the effect of noise on the estimation error.
Sturm \textit{et al.} \cite{sturm2013spectrally} considered the multi-user 
multi-input multi-output (MU-MIMO) radar and used equally spaced subcarriers per 
user to achieve full bandwidth sensing resolution.
However, this method has the disadvantage of reduced maximum unambiguous distance. 
To overcome this disadvantage,
Hakobyan \textit{et al.} \cite{hakobyan2016novel} used 
non-equidistant dynamic subcarriers instead of equally spaced subcarriers. 
Furthermore, an optimized subcarrier interleaving (SI) 
scheme is proposed to 
maximize sidelobes reduction 
while addressing the deteriorating effect 
of dynamic subcarriers 
on sidelobes. 
However, this approach cannot overcome the deterioration of sidelobes in velocity estimation. 
To this end, Knill \textit{et al.} \cite{knill2018high} presented frequency-agile sparse OFDM radar and applied a two-dimensional (2D) compressed sensing (CS) method 
to obtain the complete range-velocity profile. 
This method can be applied in the sparse time-frequency domain, but not flexibly to adapt the time-varying unlicensed spectrum bands.
To this end,
Wei \textit{et al.} \cite{wei2022robust} presented a ISAC processing method under the dynamic change of available subcarriers. Specifically, the configuration information of 
the resource block group (RBG) 
in 5G new radio (5G NR) is extracted to obtain the spectrum judgment sequence. 
This sequence is then used to zero the non-data parts of the channel information matrix and perform 
a 2D fast Fourier transform (2D FFT) to estimate range and velocity of target.
Simulation results show that the improved 2D FFT algorithm enhances the 
anti-noise performance of target sensing on non-continuous spectrum bands.
However, it doesn't solve the problem of deterioration of sidelobes.
Overall, it is obvious that the above studies provide references in terms of ISAC signal processing under non-continuous spectrum bands. For example, CS-based sensing methods are more suitable for sensing under non-continuous spectrum due to the tolerance of signal processing in undersampling. Utilizing the RBG information update processing method in 5G NR is more suitable with dynamically changing systems.
However, the above method is not applicable to the sensing processing of ISAC systems under both non-continuous spectrum and time-varying time-frequency resource allocation. Therefore, the ISAC signal processing with the capabilities of anti-noise and low sidelobes in dynamic unlicensed spectrum bands is still challenging.

In this paper, we propose a high-precision CS-based ISAC signal processing 
algorithm combined with machine learning to improve sensing performance.
This study focuses on mobile communication system, adopting OFDM as the ISAC signal,
which is the standard signal 
of mobile communication system. 
The OFDM ISAC signals in the non-continuous spectrum bands 
are collectively referred to as non-continuous OFDM (NC-OFDM) signals. 
To ensure mutual non-interference between users and optimal allocation of resources, the spectrum occupancy information is mapped in the RBG resource block, and the spectrum occupancy can be obtained by simple operations such as unmapping and demodulation \cite{3gpp2018nr,de2023survey}.
Therefore, we extract the spectrum occupancy sequence from the RBG configuration 
information in 5G NR and partially zero the channel information matrix according to 
the processing method in \cite{wei2022robust}.  
Then, we apply the proposed CS-based 2D FFT algorithm in parameter estimation on the processed channel information matrix. This algorithm is combined with machine learning techniques to search for optimal parameters. Ultimately, we obtain a power spectrum with minimal sidelobes and an enhanced signal-to-noise ratio (SNR).
It is worth noting that there are existing CS methods such as off-grid, on-grid, and one to three dimensional \cite{wu2022joint}. However, the CS-based sensing method proposed in this paper includes modeling and parameter optimization of the CS problem under dynamic non-continuous spectrum band, which distinguishes it from the existing CS methods.
The main contributions of this paper are summarized as follows.

\begin{itemize}
	\item[$\bullet$] \textbf{CS-based high-precision sensing algorithm}:
Combining the spectrum occupancy sequence information and CS theory, 
 we transform the target estimation problem into a power spectrum reconstruction 
 problem and use fast iterative shrinkage-thresholding algorithm (FISTA) \cite{beck2009fast} to solve it. Thus, high-precision sensing is achieved under dynamic,
 stochastic, and non-continuous unlicensed spectrum bands.
\end{itemize}
\begin{itemize}
	\item[$\bullet$] \textbf{Machine learning-based parameter optimization}: 
The K-fold cross validation (KCV) is applied to select the optimal regularization 
parameter $\lambda$ to realize faster convergence speed and a smaller reconstruction error
compared with the traditional methods without machine learning. 
\end{itemize}
\begin{itemize}
	\item[$\bullet$] \textbf{Algorithm superiority}: 
    Simulation results show that the proposed CS-based ISAC signal processing algorithm 
    has a higher peak-to-side lobe ratio (PSLR) and better anti-noise performance in non-continuous spectrum bands than the conventional sensing algorithm. 
    We provide the optimal regularization parameter for the SNR between 0 dB and 10 dB, 
    which results in a power spectrum with zero sidelobes.
\end{itemize}

The rest of this paper is arranged as follows. 
Section \ref{se2} introduces the NC-OFDM-based ISAC signal model and CS theory. 
Section \ref{se3} presents the ISAC signal processing over unlicensed spectrum bands. 
Section \ref{se4} reveals the optimal choice of the regular term $\lambda$ and 
the sensing performance analysis. 
Section \ref{se5} shows simulation results demonstrating improved anti-noise 
performance and PSLR of target estimation under 
non-continuous spectrum bands using the proposed algorithm. 
Finally, this paper is summarized in Section \ref{sc6}.

\textit{Notations:} $\{\cdot\}$ typically stands for a set of various index values. 
Black bold letters represent matrices or vectors.
$\mathbb{C}^{\Omega \times \mho}$ denotes the set of complex $\Omega \times \mho$ matrix. 
$\left[\cdot\right]^{\text{T}}$ and $\left(\cdot\right)^{-1}$ stand for the transpose operator and inverse operator, respectively. 
$\circ$ and $\otimes$ are the Hadamard and Kronecker product, respectively. $\|\cdot\|_0 $, $\|\cdot\|_1$, and $\|\cdot\|_2 $ are the $\ell_0$, $\ell_1$, and $\ell_2$-norm, respectively. $|\cdot|$ is absolute value.

\section{ Signal Model and Compressed Sensing}\label{se2}

\subsection{Signal Model}\label{2sub2}
The ISAC signals based on NC-OFDM and continuous OFDM have similar expression form. Assuming that the total number of subcarriers in the NC-OFDM system is $ N_\text{c}$ and the set of available subcarrier numbers obtained with the assistance of the spectrum sensing module is $ \Omega \subseteq \{1,2,\cdots,N_\text{c}\}$, the baseband transmit signal at the transmitter (TX) is expressed as \cite{5776640}
\begin{equation}\label{eq5}
	\begin{aligned}
	   x\left(t \right)_{\text{\tiny NC-OFDM}}=& \sum \limits_{m=1}^{M_{\text{sym}}}\sum \limits_{n=1}^{N_\text{c}} d_{m,n}  \exp{\left( j 2 \pi f_n t \right) } \\
	   & \times \operatorname{rect}\left( \frac{t-mT_{\text{sym}}+T_{\text{sym}}}{T_{\text{sym}}} \right),
	\end{aligned}
\end{equation}
where $ m \in \{1,2,\cdots,M_{\text{sym}} \} $ represents the index value of an NC-OFDM symbol in total $ M_{\text{sym}} $ NC-OFDM symbols, and $ n \in \Omega $ represents the sequence number of an available subcarrier in total $ N_c $ subcarriers, $ T_{\text{sym}} =T_{\text{ofdm}} +T_{\text{cp}} $ is the total duration of the NC-OFDM symbol, $T_{\text{ofdm}}$ is the elementary NC-OFDM symbol duration, $T_{\text{cp}}$ is the length of cyclic prefix (CP). $ d_{m,n} $ is the modulation symbol at the $ m $-th NC-OFDM symbol of the $ n $-th subcarrier. $ f_n $ is the frequency of $ n $-th subcarrier, which is $ n $ times of $ \Delta f $ with $ \Delta f $ representing the subcarrier spacing  \cite{huang2017nc,5776640}.

The ISAC signal based on NC-OFDM is reflected on the surface of the target with a range of $R$ and a Doppler frequency shift $ f_\text{D} = 2v_0 f_\text{c}/ c $ is generated due to relative motion, where $ v_0$ represents the radial velocity between the target and TX, $c$ is the speed of light, and $f_\text{c}$ represents the carrier frequency. Then,  the received signal at the receiver (RX) is expressed as \cite{weiiter,5776640}
\begin{equation}\label{eq6}
   \begin{aligned}
	y(t)_{\text{\tiny NC-OFDM}} = &\sum_{m=1}^{M_{\text{sym}}} \sum_{n=1}^{N_\text{c}} \alpha_{m, n} d_{m, n}  \exp \left(j 2 \pi f_n(t-\tau)\right) \\
	& \times \exp \left(j 2 \pi f_\text{D} t\right)  \operatorname{rect}\left(\frac{t-m T_{\text{sym}}+T_{\text{sym}}-\tau}{T_{\text{sym}}}\right),
\end{aligned}
\end{equation}
where $\alpha _{m,n}$ represents the channel coefficient at the $ m $-th OFDM symbol of the
$n$-th subcarrier. $\tau =2R/c$ is the time delay \cite{9737357}. In \cite{5776640}, the improved signal processing algorithm based on the radar signal processing algorithm in the symbol modulation domain was proposed. In order to highlight the influence of delay and Doppler frequency shift of target, \eqref{eq6} is rewritten as \cite{5776640}
\begin{equation}\label{eq7}
	\begin{aligned}
		y(t)_{\text{\tiny NC-OFDM}}= & \sum_{m=1}^{M_{\text{sym}}} \exp \left(j 2 \pi f_\text{D} t\right) \sum_{n=1}^{N_\text{c}} \alpha_{m, n} \\
		& \times \left\{d_{m, n} \exp \left(-j 2 \pi f_n \frac{2 R}{c}\right)\right\} \\
		& \times \exp \left(j 2 \pi f_n t\right)  \operatorname{rect}\left(\frac{t-m T_{\text{sym}}+T_{\text{sym}}-\frac{2 R}{c}}{T_{\text{sym}}}\right).
	\end{aligned}
\end{equation}

\begin{figure*}[!ht]
	\centering
    \includegraphics[width=0.8\textwidth]{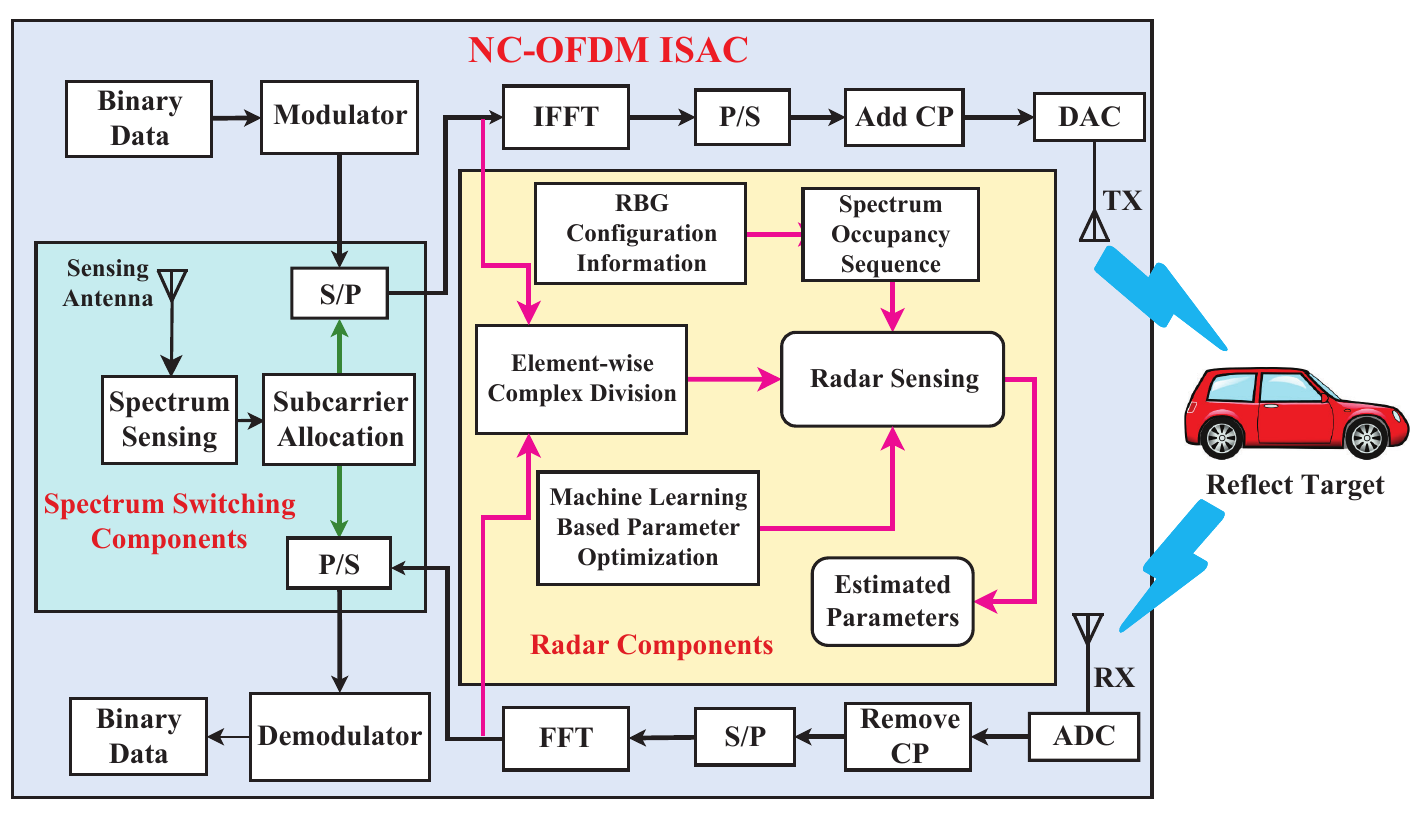}
    \caption{ Structure of NC-OFDM ISAC system, with abbreviations CP: cyclic prefix, FFT: fast Fourier transforms, IFFT: inverse fast Fourier transform, ADC: analog-to-digital converter, DAC: digital-to-analog converter, S/P: serial-to-parallel converter, P/S: parallel-to-serial converter.}
    \label{nc structure}
\end{figure*}

As shown in Fig \ref{nc structure}, the spectrum occupancy is dynamically detected and the available subcarriers are allocated to unlicensed users. At the radar component, the spectrum occupancy information in the current $N_\text{c}$ subcarriers can be obtained through the RBG configuration information module in ISAC system based on NC-OFDM. The output result of spectrum occupancy sequence module can be represented by a spectrum occupancy sequence $\textbf{A}_m \in \mathbb{C}^{N_\text{c} \times 1}$, where $ \textbf{A}_m=\left[ a_{1,m},a_{2,m},\cdots,a_{N_\text{c},m} \right]^{\mathrm{T}}$ and $[\cdot]^{\text{T}}$ represents transpose operator. $a_{i,m}=0$ indicates that the $i$-th subcarrier of the $m$-th symbol is unavailable, and $a_{i,m}=1$ indicates that the $i$-th subcarrier of the $m$-th symbol is available. The number of non-zero elements in $\textbf{A}_m$ is $\mathcal{N}_m$, and $\mathcal{N}_m$ is the number of subcarriers occupied at the $m$-th NC-OFDM symbol. Finally, the radar sensing module integrates information from the element-wise complex division module, the spectrum occupancy sequence module, and the machine learning-based parameter optimization module to estimate the range and velocity information of target carried by the reflected signal. It is worth mentioning that the compatibility between physical layer and upper layers needs to be considered in ISAC system.

The spectrum occupancy of NC-OFDM ISAC system is dynamic. Especially, ISAC system works in the unlicensed spectrum bands and the spectrum switching will be performed according to the spectrum switching component. The following research is carried out in two scenarios \cite{8275896,liuyang}. It is crucial to note that the band for unlicensed users is called the available band, while the band for licensed users is the unavailable band.

\begin{figure}[!ht]
	\centering
	\includegraphics[width=0.4\textwidth]{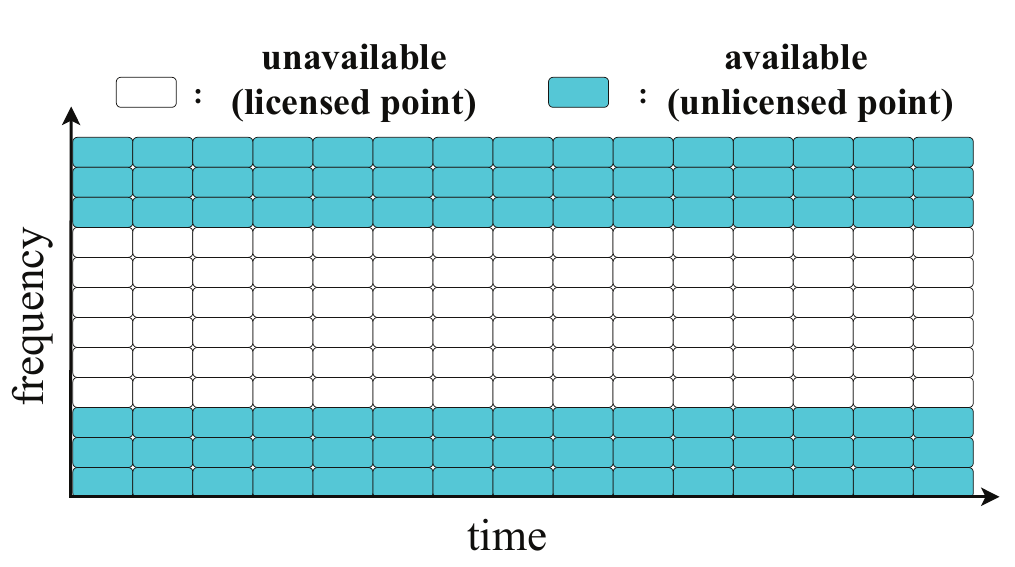}
	\caption{Spectrum occupancy under scenario 1.}
	\label{scenario 1}
\end{figure}

\begin{figure}[!h]
	\centering
	\includegraphics[width=0.4\textwidth]{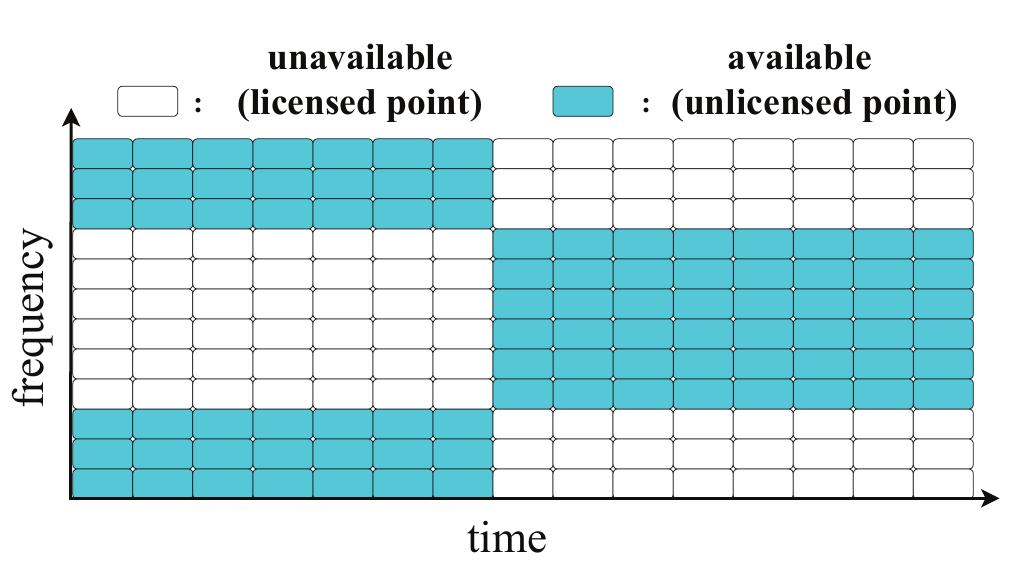}
	\caption{Spectrum occupancy under scenario 2.}
	\label{scenario 2}
\end{figure}

 \begin{itemize}
 	\item[$\bullet$] \textbf{Scenario 1}:
As shown in Fig. \ref{scenario 1}, the spectrum bands occupied by licensed users and unlicensed users are not switched within the target sensing period. 
        \item[$\bullet$] \textbf{Scenario 2}:
As shown in Fig. \ref{scenario 2}, the spectrum bands occupied by licensed users and unlicensed users are switched within the target sensing period.
 \end{itemize}
 
 Given that the study of this paper employs the CS theory, in order to facilitate the understanding of Section \ref{se3}, the CS theory is introduced in Section \ref{2sub1}.
 

\subsection{Compressed Sensing Theory}\label{2sub1}

CS is a technique for finding sparse solutions to overdetermined linear systems. It allows the recovery of the original signal with high accuracy using a small amount of undersampling information by solving an optimization problem with a measurement matrix orthogonal to the transformation matrix \cite{he2011compressive}.

Under an orthogonal basis $ \psi \in \mathbb{C}^{N \times N} $, the transformation of signal $ \textbf{x} $ is $ \mathbf{\theta} = (\mathbf{\psi})^{\mathrm{T}}\textbf{x} $. If the number of non-zero elements in the sparse vector $\theta$ is equal to an integer $ K $, the sparsity of the signal $ \textbf{x} $ is called $ K $-sparse. CS is applied to solve the problem of reconstructing the original vector $\textbf{x} $ from the observed vector $\textbf{y}$, which is given by~\cite{he2011compressive}
\begin{equation}\label{eq1}
	\textbf{y}=\mathbf{\Phi}\textbf{x}= \textbf{A}\theta,
\end{equation}
where $ \textbf{y} \in \mathbb{C}^{M \times 1} (M < N)$, $ \textbf{x} \in \mathbb{C}^{N \times 1} $, $ \mathbf{\Phi} \in \mathbb{C}^{M \times N} $ is a measurement matrix, and $ \textbf{A} \in \mathbb{C}^{M \times N} = \mathbf{\Phi} \mathbf{\psi}$ is called sensing matrix. Solving \eqref{eq1} directly is an ill-conditioned problem. If $ M\gg K $, and the sensing matrix $ \textbf{A} $ satisfies the restricted isometry property (RIP) condition \cite{he2011compressive}, the problem of \eqref{eq1} can be transformed into the minimum $ \ell_0 $ norm problem, namely~\cite{hochba1997approximation} 
\begin{equation}\label{eq2}
   \min \limits_{\theta} \Vert \theta \Vert_0 \quad  \mathrm{s.t.} \quad \textbf{A}\theta = \textbf{y},
\end{equation}
where $\| \cdot \|_0$ represents the $ \ell_0 $-norm.

The above problem is non-deterministic polynomial-time hard (NP-hard) \cite{hochba1997approximation}. There are several types of convex optimization algorithms, combinatorial algorithms and greedy algorithms regarding the solution of the above problem. Among them, convex optimization algorithms are able to find the global optimal solution and are more stable~\cite{kutyniok2013theory}. \cite{candes2008introduction} mentions that the solution of the minimum $ \ell_0 $ norm problem is equivalent to the solution of the minimum $ \ell_1 $ norm problem according to the sparsity of reconstructed signal and the correlation of sensing matrix, namely, the non-convex optimization problem is transformed into a convex optimization problem. Then, \eqref{eq2} is transformed into~\cite{hochba1997approximation}
\begin{equation}\label{eq3}
	\min \limits_{\theta} \Vert \theta \Vert_1 \quad \mathrm{s.t.} \quad \textbf{A}\theta =\textbf{y},
\end{equation}
where $\| \cdot \|_1$ represents $\ell_1$-norm.

In practice, the observed vector $\textbf{y}$ is generally contains noise and \eqref{eq1} is transformed into $ \textbf{y}= \mathbf{\Phi}\textbf{x}+ \varepsilon $, where $\varepsilon$ is a additive white Gaussian noise (AWGN). Therefore, \eqref{eq3} is transformed into \cite{taubock2008compressed}
\begin{equation}\label{eq4}
	\min \limits_{\theta} \Vert \theta\Vert_1 \quad \mathrm{s.t.} \quad \Vert \textbf{A}\theta-\textbf{y} \Vert_2  \leqslant \varepsilon.
\end{equation}
By solving the above convex optimization problem, we achieve the recovery of the original data from the undersampling data.
Regarding the solution of convex optimization problems in compressed sensing, there are a large number of algorithms that can be implemented, include basis pursuit (BP) \cite{huang2017nc}, least absolute sampling and selection algorithm (LASSO) \cite{han2013compressive} and FISTA \cite{beck2009fast}.
Meanwhile, the focus of this paper is not on the optimization of the CS reconstruction algorithm, but on the approach to transform the sensing signal model of this paper into a CS reconstruction model, which will be solved in Section~\ref{se3}.

\section{ISAC Signal Processing over Unlicensed Spectrum Bands}\label{se3}
In this section, the ISAC signal processing algorithm over unlicensed spectrum bands is proposed. Specifically, we first describe how to obtain the channel information matrix. Then the sensing algorithms for range and radial velocity of the target are presented, respectively.
We denote this ISAC signal processing method as ``Joint CS and Machine Learning ISAC Signal Processing Algorithm'', abbreviated as JCMSA.

In 2011, Sturm \textit{et al.} \cite{5776640} proposed the 2D FFT method, which utilizes a 2D Fourier transform to estimate the velocity and range of target. Velocity and range estimations can be obtained by the modulation symbols sequence $\mathbf{d}_{\text{TX}}\left(m,n' \right) $ and $\mathbf{d}_{\text{RX}}\left(m,n' \right) $, where $n' \in \left\{ 1,2,\cdots, N_{\text{c}} \right\}$ represents the $n'$-th subcarrier of the total $N_{\text{c}}$ subcarriers in NC-OFDM system. Observing (\ref{eq7}), the delay and Doppler frequency shift caused by the target are independent in the modulation symbol~\cite{weisurvey}. Thus, the modulation symbol on the RX can be expressed as \cite{5776640}
\begin{equation}\label{eq8}
	\begin{aligned}
 	\mathbf{d}_{\text{RX}}\left(m N_\text{c}, n' \right)=&\alpha_{m , n'} \mathbf{d}_{\text{TX}}\left(mN_\text{c},n'\right) \exp \left(-j 2 \pi f_n' \frac{2 R}{c}\right) \\ 
 	&\times \exp \left(j 2 \pi m T_{\text{sym}} f_\text{D}\right). 	
	\end{aligned}
\end{equation}
For a clear representation, the modulation symbols can be converted to the matrix form, where each row represents a subcarrier of the same frequency and each column represents an NC-OFDM symbol. (\ref{eq8}) can be expressed as \cite{wei2023carrier}
\begin{equation}\label{eq9}
  \left(\mathbf{D}_{\text{RX}}\right)_{m, n'}=\alpha_{m, n'}\left(\mathbf{D}_{\text{TX}}\right)_{m, n'} \circ\left(\mathbf{d}_\text{r} \otimes \mathbf{d}_\text{v}\right)_{m, n'} ,
\end{equation}
where $\circ $ denotes Hadamard product, and $\otimes$ denotes Kronecker product. Through  an element-wise complex division, the channel information matrix $(\mathbf{D}_{\text{div}})_{m,n'}$ is obtained by~\cite{wei2023carrier}
\begin{equation}\label{eq10}
	\left(\mathbf{D}_{\text{div }}\right)_{m, n'}=\frac{\left(\mathbf{D}_{\text{Rx}}\right)_{m, n'}}{\left(\mathbf{D}_{\text{Tx}}\right)_{m, n'}}=\alpha_{m, n'}\left(\mathbf{d}_\text{r} \otimes \mathbf{d}_\text{v}\right)_{m,n'},
\end{equation}
where the vectors $\mathbf{d}_\text{r} $ and $\mathbf{d}_\text{v} $ can be transformed into~\cite{wei2023carrier}
\begin{equation}\label{eq11}
	\mathbf{d}_\text{r}(n')=\exp \left(-j 2 \pi f_n' \frac{2 R}{c}\right) , \quad n'=1, \ldots, N_\text{c}
\end{equation}
and
\begin{equation}\label{eq12}
	\mathbf{d}_\text{v}(m)=\exp \left(j 2 \pi m T_{\text{sym}} \frac{2 v_0 f_\text{c}}{c}\right). \quad m=1, \ldots, M_{\text{sym}}
\end{equation}
Finally, the range and radial velocity of target can be obtained by applying (\ref{eq11}) and (\ref{eq12}) to perform inverse discrete Fourier transform (IDFT) and discrete Fourier transform (DFT) \cite{5776640}.

When the ISAC-based mobile communication systems work on unlicensed spectrum bands with spectrum occupancy as shown in Fig. \ref{scenario 1}  and Fig. \ref{scenario 2}, the 2D FFT method shows deterioration in sensing performance. Specifically, the spectrum holes caused by the non-continuous spectrum bands can lead to deterioration of the Fourier sidelobes of the power spectrum and degradation of anti-noise performance \cite{11}. To address the above problems, we proposed a novel algorithm, namely JCMSA, which transforms the power spectrum estimation problem into a compressive reconstruction problem, thereby obtaining lower sidelobes and better anti-noise performance.

\subsection{Range Estimation Algorithm}\label{3sub1}
First, we use modulation symbol division to obtain the unprocessed channel information matrix. Due to the non-continuous spectrum, the noise power of the unavailable points in the unprocessed channel information matrix is amplified \cite{wei2022robust}. To reduce the noise, we first extract the RBG configuration information in 5G NR to obtain the spectrum occupancy sequence $\mathbf{A}_{m}$. Then, we combine $\mathbf{A}_{m}$ and the unprocessed channel information matrix to set the unavailable points to zero. The processed channel information matrix $\mathbf{D}_{\text{R1}}$ for scenario 1 is expressed as (\ref{eq13}).

In (\ref{eq13}),
\begin{figure*}
  \tiny{
  \begin{align}
    \label{eq13}
	\mathbf{D}_{\text{R1}}=\left[ \begin{array}{cccc}
	s_{1,1}+\varepsilon_{1,1} & s_{1,2}+\varepsilon_{1,2} & \cdots & s_{1,M_{\text{sym}}}+\varepsilon_{1,M_{\text{sym}}} \\ 
    s_{2,1}+\varepsilon_{2,1} & s_{2,2}+\varepsilon_{2,2} & \cdots & s_{2,M_{\text{sym}}}+\varepsilon_{2,M_{\text{sym}}} \\
    \vdots & \vdots & \ddots & \vdots \\ 
    s_{\frac{\mathcal{N}_m}{2},1}+\varepsilon_{\frac{\mathcal{N}_m}{2},1} & s_{\frac{\mathcal{N}_m}{2},2}+\varepsilon_{\frac{\mathcal{N}_m}{2},2} & \cdots & s_{\frac{\mathcal{N}_m}{2},M_{\text{sym}}}+\varepsilon_{\frac{\mathcal{N}_m}{2},M_{\text{sym}}} \\ 
     0     &      0 & \cdots &     0 \\ 
     \vdots& \vdots & \ddots & \vdots \\
     0     &     0  & \cdots & 0 \\
    s_{N_\text{c}-\frac{\mathcal{N}_m}{2}+1,1}+\varepsilon_{N_\text{c}-\frac{\mathcal{N}_m}{2}+1,1} & s_{N_\text{c}-\frac{\mathcal{N}_m}{2}+1,2}+\varepsilon_{N_\text{c}-\frac{\mathcal{N}_m}{2}+1,2} & \cdots & s_{N_\text{c}-\frac{\mathcal{N}_m}{2}+1,M_{\text{sym}}}+\varepsilon_{N_\text{c}-\frac{\mathcal{N}_m}{2}+1,M_{\text{sym}}} \\  
    \vdots & \vdots & \ddots & \vdots \\ 
  s_{N_\text{c},1}+\varepsilon_{N_\text{c},1} & s_{N_\text{c},2}+\varepsilon_{N_\text{c},2} & \cdots & s_{N_\text{c},M_{\text{sym}}}+\varepsilon_{N_\text{c},M_{\text{sym}}}
	\end{array}\right],
   \end{align}}
   {\noindent} \rule[-10pt]{18cm}{0.1em}
\end{figure*}
$s_{m,n'}$ and $\varepsilon_{m,n'}$ denote that the data and noise after the modulation symbols have undergone element-wise complex division, respectively. The position of the zero-row vector in $ \mathbf{D}_{\text{R1}}$ corresponds to the spectrum occupancy sequence $\textbf{A}_m$. From (\ref{eq10}), it is revealed that the linear phase shift of range is only along the frequency axis. Therefore, by processing the columns of $ \mathbf{D}_{\text{R1}}$, the range of target information can be obtained.

\subsubsection{Compressive reconstruction of the power spectrum of range}
$ \mathbf{D}_{\text{R1}}$ is split into $M_{\text{sym}}$ column vectors, with expressions as follows.  
\begin{equation}\label{eq14}
	\mathbf{D}_{\text{R1}}= \left( \mathbf{y}_1 , \mathbf{y}_2 , \cdots, \mathbf{y}_m, \cdots , \mathbf{y}_{M_{\text{sym}}}\right) ,
\end{equation}
where $\mathbf{y}_m \in \mathbb{C}^{N_\text{c} \times 1}$ is the $m$-th column vector of $\mathbf{D}_{\text{R1}}$. According to (\ref{eq13}), the position of the zero element in $\mathbf{y}_m$ corresponds to the zero element in $\mathbf{A}_m$, which is expressed by 
\begin{equation}\label{eq15}
	\mathbf{y}_{m} = \mathbf{A}_m \circ \mathbf{y}_{m(\text{full})},
\end{equation}
where $\mathbf{y}_{m(\text{full})}$ represents the $m$-th column vector of the channel information matrix obtained from the OFDM signal that occupies all subcarriers. According to 2D FFT algorithm, the power spectrum of the range can be obtained by performing IDFT on the column vectors of the channel information matrix. Hence, the above operation can be expressed by
\begin{equation}\label{eq16}
	\mathbf{\Psi} \mathbf{y}_{m(\text{full})}=\mathbf{P}_m,
\end{equation}
where $\mathbf{\Psi} \in \mathbb{C}^{N_\text{c} \times N_\text{c}} $ is IDFT matrix, and $\mathbf{P}_m \in \mathbb{C}^{N_\text{c} \times 1}$ represents the power spectrum of range. Given that $\mathbf{\Psi}$ is an invertible square matrix, (\ref{eq16}) can be transformed as
\begin{equation}\label{eq17}
	 \mathbf{y}_{m(\text{full})}=\mathbf{\Psi}^{-1} \mathbf{P}_m,
\end{equation}
where $(\cdot)^{-1}$ is the inverse operator.

\begin{figure}[!ht]
	\centering
	\includegraphics[width=0.45\textwidth]{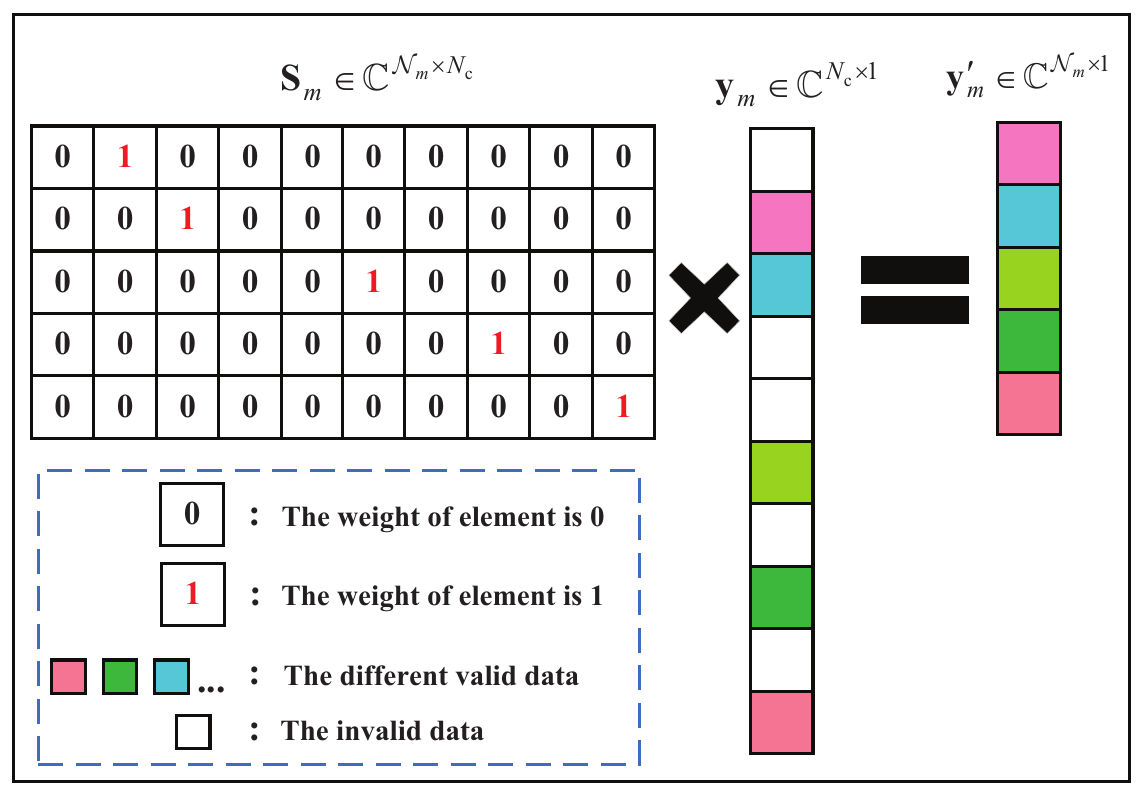}
	\caption{The process of selecting valid data.}
	\label{selection matrix work}
\end{figure}

Substituting (\ref{eq17}) into (\ref{eq15}), and expanding $\mathbf{A}_m $ into the matrix $\mathbf{A'}_m 
\in \mathbb{C}^{N_\text{c} \times N_\text{c}}$ for Hadamard product operation with $\mathbf{\Psi}$, we have
\begin{equation}\label{eq18}
	\mathbf{y}_m=\mathbf{A}_m \circ \left( \mathbf{\Psi}^{-1}\mathbf{P}_m\right)= \left(\mathbf{A'}_m \circ \mathbf{\Psi}^{-1} \right) \mathbf{P}_m,
\end{equation}
where $\mathbf{A'}_m = \left(\mathbf{A}_m,\mathbf{A}_m,\cdots ,\mathbf{A}_m \right) $. Define a selection matrix $\mathbf{S}_m \in \mathbb{C}^{\mathcal{N}_m \times N_\text{c}}$ to filter out the valid data on the $m$-th NC-OFDM symbol, and the intuitive filtering process is shown in Fig. \ref{selection matrix work}. Multiplying both sides of (\ref{eq18}) by $\mathbf{S}_m $, we can get
\begin{equation}\label{eq19}
	\mathbf{y'}_m =\mathbf{J}_m \mathbf{P}_m,
\end{equation}
where 
\begin{equation}\label{eq20}
	\begin{cases}
		\mathbf{y'}_m = \mathbf{S}_m\mathbf{y}_m \\
		\mathbf{J}_m \in \mathbb{C}^{\mathcal{N}_m\times N_{\text{c}}} = \mathbf{S}_m \left(\mathbf{A'}_m \circ \mathbf{\Psi}^{-1} \right)
	\end{cases}
\end{equation}

The problem of solving (\ref{eq19}) is equivalent to (\ref{eq1}). According to the CS theory in section \ref{2sub1}, the above problem can be solved if the following two conditions are satisfied. 

\textbf{Condition 1:} $\mathbf{P}_m$ is a sparse vector.

\textbf{Condition 2:} The sensing matrix $\mathbf{J}_m$ satisfies RIP. 

Due to the limited number of targets in radar sensing, one target corresponds to a peak in range-velocity profile, so that radar signals are sparse in range-velocity profile space \cite{knill2018high}. Therefore, $\mathbf{P}_m$ is a sparse vector. According to (\ref{eq20}), it is revealed that $\mathbf{J}_m$ is a partial DFT matrix. It is shown in \cite{candes2006near} that the partial DFT matrix satisfies the RIP. Thus, (\ref{eq19}) can satisfy the condition for accurate reconstruction of the spare vector $\mathbf{P}_m$. The received signal usually contains AWGN, where the sparse power spectrum can be obtained by solving (\ref{eq4}). In this paper, the FISTA \cite{beck2009fast} is used to undergo the solution, and the power spectrum $\mathbf{P}_m$ of range is finally obtained after several iterations by choosing a suitable regularization parameter $\lambda$. Details about the FISTA and parameter selection are given in section \ref{se4}.
\subsubsection{Range estimation process} With the following four steps, the unprocessed channel information matrix can be used to estimate the target's range information.
\begin{itemize}
    \item \textbf{Step 1:} The processed channel information matrix (e.g., Eq. (\ref{eq13})) is obtained by modulation symbol division and the spectrum occupancy sequence $\mathbf{A}_m$. 
    \item \textbf{Step 2:} For each non-zero column vector $\mathbf{y}_m $ of the processed channel information matrix, the FISTA is used for solving (\ref{eq19}) to obtain $\mathbf{P}_m$.
    \item \textbf{Step 3:} To improve anti-noise performance, we accumulate and normalize $M_{\text{a}}$ range power spectra, where $M_{\text{a}} \le M_{\text{sym}}$ represents the number of non-zero column vectors $ \mathbf{y}_m$. Given that the velocity of target is unknown and the phase of $M_{\text{a}}$ range power spectra are not aligned, the accumulation is non-coherent.
    \item \textbf{Step 4:} The peak of power spectrum is searched to obtain the peak index value $k_0$, which is substituted into (\ref{eq21}) to get the range of target.
\end{itemize}
\begin{equation}\label{eq21}
	\hat{R}=\dfrac{c}{2N_\text{c}\Delta f}(k_0-1).
\end{equation}
The improved range estimation algorithm is given in \hyperref[tab1]{Algorithm 1}, where $\mathbf{D'}_{\mathrm{R}}( :,m)$ denotes the elements on each row and the $m$-th column of $\mathbf{D'}_{\mathrm{R}}$.

In terms of target range estimation, the algorithm proposed in this paper is suitable for the scenario where the spectrum bands are dynamically changed and only the spectrum occupancy at each moment needs to be known. Therefore, this algorithm is highly adaptable.

\begin{table}[!ht]
\centering
\label{tab1}
\resizebox{0.45\textwidth}{!}{
\setlength{\arrayrulewidth}{1.3pt}
\begin{tabular}{rllll}
\hline
\multicolumn{5}{l}{\textbf{Algorithm 1:} Improved Range Estimation Algorithm}  \\ \hline
\multirow{-7}{*}{\textbf{Input:} }            & \multicolumn{4}{l}{\begin{tabular}[c]{@{}l@{}}Unprocessed channel information matrix $\mathbf{D}_{\mathrm{R}}$;\\ Spectrum occupancy sequence $\mathbf{A}_m$, $m \in \left \{  {1,2,\cdots,M_{\text{sym}}}\right \} $;\\ Selection matrix $\mathbf{S}_m$; \\ The sensing matrix $\mathbf{J}_m$ for compressive reconstruction;\\ Regularization parameter $\lambda$;\\ The number of NC-OFDM symbols $M_{\text{sym}}$;\\ The total number of subcarriers $N_{\text{c}}$;\end{tabular}} \\
\textbf{Output:}         & \multicolumn{4}{l}{The range of target $\hat{R}$.} \\ 
\textbf{Part 1:}         & \multicolumn{4}{l}{} \\ 
1:                       & \multicolumn{4}{l}{Initialize the processed channel information matrix $\mathbf{D'}_{\mathrm{R}} \in \mathbb{C}^{N_{\text{c}}\times M_{\text{sym}}}$;}  \\
2:                       & \multicolumn{4}{l}{$\textbf{For}$ $\mathbf{D}_{\mathrm{R}}$ $m$ in $M_{\text{sym}}$ $\textbf{do}$}  \\
\multirow{-2}{*}{3:}                        & \multicolumn{4}{l}{\begin{tabular}[c]{@{}l@{}}$\hspace{1em}$ The $m$-th column vector of $\mathbf{D}_{\mathrm{R}}$ and $\mathbf{A}_m$ are performed \\ $\hspace{1em}$ the Hadamard product;\end{tabular}}  \\
4:                       & \multicolumn{4}{l}{$\hspace{1em}$ Results of the 3-th step are assigned to $ \mathbf{D'}_{\mathrm{R}}( :,m)$;}  \\
5:                       & \multicolumn{4}{l}{$\textbf{End}$ $\textbf{For}$}     \\
6:                       & \multicolumn{4}{l}{Obtain $\mathbf{D'}_{\mathrm{R}}$;}  \\
\textbf{Part 2:}         & \multicolumn{4}{l}{} \\ 
7:                       & \multicolumn{4}{l}{Initialize a range power spectrum $\mathbf{P}\in \mathbb{C}^{1 \times N_{\text{c}}}$ and $i=1$;}   \\
8:                       & \multicolumn{4}{l}{$\textbf{For}$ $\mathbf{D'}_{\mathrm{R}}$ $m$ in $M_{\text{sym}}$ $\textbf{do}$}  \\
9:                       & \multicolumn{4}{l}{$\hspace{1em}$ $\textbf{If}$ the $m$-th column vector of $\mathbf{D'}_{\mathrm{R}}$ $>0$ $\textbf{do}$}    \\
10:                      & \multicolumn{4}{l}{$\hspace{2em}$ Set the regularization parameter $\lambda$ according to Table \ref{tab_5} and  \ref{tab_6};}  \\
11:                      & \multicolumn{4}{l}{$\hspace{2em}$ Set the maximum number of iterations $I_{\text{max}}$;}  \\
12:                      & \multicolumn{4}{l}{$\hspace{2em}$ Set the threshold of iterations error $\Theta$;}  \\
\multirow{-2}{*}{13:}                        & \multicolumn{4}{l}{ \begin{tabular}[c]{@{}l@{}}$\hspace{2em}$ The $m$-th column vector of $\mathbf{D'}_{\mathrm{R}}$ and selection matrix $\mathbf{S}_m$ \\ $\hspace{2em}$ perform (\ref{eq20}) to obtain $\mathcal{y'}_m$;\end{tabular}} \\
14:      & \multicolumn{4}{l}{$\hspace{2em}$ Take $\mathcal{P}_1=\mathcal{Z}\in \mathbb{R}^{N_{\text{c}}}$, $t_1=1$;}  \\
15:                      & \multicolumn{4}{l}{$\hspace{2em}$ $\textbf{While} \ \text{error}(k+1)-\text{error}(k)\ge \Theta$ and $k \le I_{\text{max}}$ $\textbf{do}$}  \\
16:                      & \multicolumn{4}{l}{$\hspace{3em}$ $\mathcal{Z}_k = p_{L}(\mathcal{P}_k)$ according to Table \ref{tab_3};}  \\
17:                      & \multicolumn{4}{l}{$\hspace{3em}$ $t_{k+1}=\frac{1+\sqrt{1+4t_{k}^{2}}}{2}$;} \\
18:                      & \multicolumn{4}{l}{$\hspace{3em}$ $\mathcal{P}_{k+1}=\mathcal{Z}_k+\left(\frac{t_k-1}{t_{k+1}}\right)(\mathcal{Z}_k-\mathcal{Z}_{k-1})$;} \\
19:                      & \multicolumn{4}{l}{$\hspace{3em}$ $\text{error}(k+1)=\|\mathbf{J}_{m}\mathcal{Z}_k-\mathcal{y'}_{m}\|_2$;}  \\
20:                      & \multicolumn{4}{l}{$\hspace{2em}$ $\textbf{End While}$}   \\
21:                      & \multicolumn{4}{l}{$\hspace{2em}$ Obtian the reconstruction data $\mathcal{Z}$;}  \\
22:                      & \multicolumn{4}{l}{$\hspace{2em}$ $\mathbf{P}=\mathbf{P}+|\mathcal{Z}|$, where $|\cdot|$ is an absolute value;} \\
23:                      & \multicolumn{4}{l}{$\hspace{2em}$ $i=i+1$;}  \\
24:                      & \multicolumn{4}{l}{$\hspace{1em}$ $\textbf{End If}$}  \\
25:                      & \multicolumn{4}{l}{$\textbf{End For}$}  \\
26:                      & \multicolumn{4}{l}{$\mathbf{P}=\mathbf{P} / i$;} \\
\textbf{Part 3:}         & \multicolumn{4}{l}{} \\ 
27:                      & \multicolumn{4}{l}{Search for the peak value of $\mathbf{P}$ and get the peak index value;}\\
28:                      & \multicolumn{4}{l}{Substitute the index value into (\ref{eq21});}  \\
\hline
\end{tabular}}
\end{table}

\subsection{Velocity Estimation Algorithm}
Since the Doppler effect causes a linear phase shift along the time axis, the target's velocity can be obtained by processing row vectors of the channel information matrix.
Firstly, the processed channel information matrix need to be obtained. However, unlike the range estimation algorithm, $\mathbf{A}_m$ cannot be directly used for Hadamard product. Therefore the spectrum occupancy sequences in $M_{\text{sym}}$ symbol times need to be processed.

Firstly, the matrix $\mathbf{Z} \in \mathbb{C}^{N_\text{c} \times M_{\text{sym}}}$ is obtained by arranging the spectrum occupancy sequences $\mathbf{A}_m$ in $M_{\text{sym}}$ symbol times in rows.
\begin{equation}\label{eq22}
	\mathbf{Z} =\left( \mathbf{A}_1, \mathbf{A}_2, \cdots , \mathbf{A}_m, \cdots , \mathbf{A}_{M_{\text{sym}}} \right).
\end{equation}
Then, $\mathbf{Z}$ is Hadamard product with the corresponding unprocessed channel information matrix to obtain the processed channel information matrix. The processed channel information matrix $\mathbf{D}_{\mathrm{V2}}$ under scenario 2 is expressed as (\ref{eq23}),
\begin{figure*}	
                \begin{equation}{ \fontsize{4}{6}
                \begin{aligned} 
				 & \mathbf{D}_{\mathrm{V2}}  = \\ &
				 \left[ 
                \setlength{\arraycolsep}{0.5pt}
                \renewcommand{\arraystretch}{2.2}
				\begin{array}{cccccc}
					q_{1,1}+\varepsilon_{1,1}& \cdots & q_{1,\frac{M_{\text{sym}}}{2}}+\varepsilon_{1,\frac{M_{\text{sym}}}{2}} & 0 & \cdots & 0 \\
					\vdots	& \ddots & \vdots & \vdots & \ddots & \vdots  \\
					q_{\frac{\mathcal{N}_m}{2},1}+\varepsilon_{\frac{\mathcal{N}_m}{2},1}& \cdots & q_{\frac{\mathcal{N}_m}{2},\frac{M_{\text{sym}}}{2}}+\varepsilon_{\frac{\mathcal{N}_m}{2},\frac{M_{\text{sym}}}{2}}& 0 & \cdots & 0 \\
					0	& \cdots & 0 & q_{\frac{\mathcal{N}_m}{2}+1,\frac{M_{\text{sym}}}{2}+1}+\varepsilon_{\frac{\mathcal{N}_m}{2}+1,\frac{M_{\text{sym}}}{2}+1} & \cdots & q_{\frac{\mathcal{N}_m}{2}+1,M_{\text{sym}}}+\varepsilon_{\frac{\mathcal{N}_m}{2}+1,M_{\text{sym}}} \\
					\vdots	& \ddots & \vdots & \vdots & \ddots & \vdots  \\
					0 	& \cdots  & 0  & q_{N_\text{c}-\frac{\mathcal{N}_m}{2},\frac{M_{\text{sym}}}{2}+1}+\varepsilon_{N_\text{c}-\frac{\mathcal{N}_m}{2},\frac{M_{\text{sym}}}{2}+1} & \cdots & q_{N_\text{c}-\frac{\mathcal{N}_m}{2},M_{\text{sym}}}+\varepsilon_{N_\text{c}-\frac{\mathcal{N}_m}{2},M_{\text{sym}}}\\
					q_{N_\text{c}-\frac{\mathcal{N}_m}{2}+1,1}+\varepsilon_{N_\text{c}-\frac{\mathcal{N}_m}{2}+1,1}& \cdots & q_{N_\text{c}-\frac{\mathcal{N}_m}{2}+1,\frac{M_{\text{sym}}}{2}}+\varepsilon_{N_\text{c}-\frac{\mathcal{N}_m}{2}+1,\frac{M_{\text{sym}}}{2}} & 0 & \cdots & 0 \\
					\vdots	& \ddots & \vdots & \vdots & \ddots & \vdots  \\
					q_{N_\text{c},1}+\varepsilon_{N_\text{c},1}	& \cdots & q_{N_\text{c},\frac{M_{\text{sym}}}{2}}+\varepsilon_{N_\text{c},\frac{M_{\text{sym}}}{2}} & 0 & \cdots & 0 
				\end{array} \right].
				\label{eq23}
			\end{aligned} }
                \end{equation}
			{\noindent} \rule[-10pt]{18cm}{0.1em}
\end{figure*}
where $q_{m,n'}$ represents that the data after the modulation symbols have undergone element-wise complex division. 

\subsubsection{Compressive reconstruction of the power spectrum of velocity}
The matrix $\mathbf{Z}$ is divided into $N_\text{c}$ row vectors and the $n'$-th row vector is denoted by $\mathbf{U}_{n'}$. The same operation is used for the matrix $ \mathbf{D}_{\mathrm{V}2}$ to obtain the row vector $\mathbf{k}_{n'}$ of its $n'$-th row. The derivation of the equation is performed below, which is transformed into column vectors in order to be consistent with the derivation of the range estimation. Similar to (\ref{eq15}), (\ref{eq16}) and (\ref{eq17}), we obtain the relationship in terms of velocity
\begin{subequations}\label{24}
	\begin{align}
		&(\mathbf{k}_{n'})^{\mathrm{T}} =(\mathbf{U}_{n'})^{\mathrm{T}} \circ (\mathbf{k}_{{n'}(\text{full})})^{\mathrm{T}}, \label{eq:24a} \\
		  &\mathbf{k}_{{n'}(\text{full})} \mathbf{\Upsilon} = \mathbf{Q}_{n'},  \label{eq:24b}\\
		 &\mathbf{k}_{{n'}(\text{full})} = \mathbf{Q}_{n'} \mathbf{\Upsilon}^{-1}, \label{eq:24c}
	\end{align}
\end{subequations}
where $\mathbf{k}_{{n'}(\text{full})} $ is the $n'$-th row vector of the channel information matrix obtained from the OFDM signal that occupies all subcarriers, $ \mathbf{\Upsilon} \in \mathbb{C}^{M_{\text{sym}} \times M_{\text{sym}}} $ is DFT matrix, and $\mathbf{Q}_{n'} \in \mathbb{C}^{1 \times M_{\text{sym}}}$ represents the power spectrum of velocity. 
Substituting (\ref{eq:24c}) into (\ref{eq:24a}), and expanding $\mathbf{U}_{n'}$ into matrix $\mathbf{U'}_{n'} \in \mathbb{C}^{M_{\text{sym}} \times M_{\text{sym}}}$ for Hadamard product operation with $\mathbf{\Upsilon}$, we can get
\begin{equation}\label{eq25}
	\begin{aligned}
	(\mathbf{k}_{n'})^{\mathrm{T}} & =(\mathbf{U}_{n'})^{\mathrm{T}} \circ (\mathbf{Q}_{n'} \mathbf{\Upsilon}^{-1})^{\mathrm{T}} \\
	& = (\mathbf{U}_{n'})^{\mathrm{T}} \circ \left[ (\mathbf{\Upsilon}^{-1})^{\mathrm{T}} (\mathbf{Q}_{n'})^{\mathrm{T}}\right] \\
	& = \left[(\mathbf{U'}_{n'})^{\mathrm{T}} \circ (\mathbf{\Upsilon}^{-1})^{\mathrm{T}} \right] 
	(\mathbf{Q}_{n'})^{\mathrm{T}} \\
	& = \left(\mathbf{U'}_{n'} \circ \mathbf{\Upsilon}^{-1}\right)^{\mathrm{T}}
	(\mathbf{Q}_{n'})^{\mathrm{T}},
    \end{aligned}
\end{equation}
where $ \mathbf{U'}_{n'} = \left(\mathbf{U}_{n'} , \mathbf{U}_{n'},\cdots, \mathbf{U}_{n'}\right)^{\mathrm{T}} $. Define a selection matrix $\mathbf{G}_{n'} \in \mathbb{C}^{\mathcal{M}_{n'} \times M_{\text{sym}}}$ to filter out the valid data on the $n'$-th subcarrier. $\mathcal{M}_{n'}$ is the number of valid data in the $n'$-th subcarrier and the intuitive filtering process is similar to Fig. \ref{selection matrix work}. Multiplying both sides of (\ref{eq25}) by $\mathbf{G}_{n'} $, we have
\begin{equation}\label{eq26}
	\mathbf{k'}_{n'} = \mathbf{F}_{n'} (\mathbf{Q}_{n'})^{\mathrm{T}},
\end{equation}
where
\begin{equation}\label{eq27}
	\begin{cases}
	 \mathbf{k'}_{n'} = \mathbf{G}_{n'}	(\mathbf{k}_{n'})^{\mathrm{T}} \\
	 \mathbf{F}_{n'} \in \mathbb{C}^{\mathcal{M}_{n'} \times M_{\text{sym}}} = \mathbf{G}_{n'} \left(\mathbf{U'}_{n'}\circ \mathbf{\Upsilon}^{-1}\right)^{\mathrm{T}}
	\end{cases}
\end{equation}

Since the sensing matrix $\mathbf{F}_{n'}$ is a partial IDFT matrix and $(\mathbf{k}_{n'})^{\mathrm{T}}$ is a sparse vector, the conditions for accurate reconstruction of the spare vector $\mathbf{Q}_{n'}$ are still satisfied. The FISTA is applied to solve the optimization problem.

\subsubsection{Velocity estimation process} With the following four steps, the unprocessed channel information matrix can be used to estimate the target's velocity information.
\begin{itemize}
    \item \textbf{Step 1:} The processed channel information matrix (e.g., Eq. (\ref{eq23})) is obtained by an element-wise complex division and the matrix $\mathbf{Z}$.
    \item \textbf{Step 2:} For each non-zero row vector $\mathbf{k}_n$ of the processed channel information matrix, the FISTA is applied to obtain $\mathbf{Q}_{n'}$.
    \item \textbf{Step 3:} To improve anti-noise performance, we accumulate and normalize $N_\text{b}$ velocity power spectra. $N_\text{b} \le N_\text{c}$ is the number of non-zero row vectors $\mathbf{k}_{n'}$. Given that the range of target is unknown and the phase of $N_\text{b}$ velocity power spectra not aligned, the accumulation is non-coherent.
    \item \textbf{Step 4:} The peak of power spectrum is searched to obtain the peak index value $l_0$, which is substituted into (\ref{eq28}) to get the velocity of target.
\end{itemize}
\begin{equation}\label{eq28}
	\hat{v_0} = \frac{c}{2M_{\text{sym}}T_{\text{sym}}f_\text{c}}(l_0-1).
\end{equation} 
The improved velocity estimation algorithm is given in \hyperref[tab2]{Algorithm 2}.

\begin{table}[!ht]
\centering
\label{tab2}
\resizebox{0.45\textwidth}{!}{
\setlength{\arrayrulewidth}{1.5pt}
\begin{tabular}{rllll}
\hline
\multicolumn{5}{l}{\textbf{Algorithm 2:} Improved Velocity Estimation Algorithm}   \\ \hline
\multirow{-7}{*}{\textbf{Input:} }               & \multicolumn{4}{l}{\begin{tabular}[c]{@{}l@{}}Unprocessed channel information matrix $\mathbf{D}_{\mathrm{V}}$;\\ Spectrum occupancy sequence $\mathbf{A}_m$,$m \in {1,2,\cdots,M_{\text{sym}}}$;\\ Selection matrix $\mathbf{G}_{n'}$; \\ The sensing matrix $\mathbf{F}_{n'}$ for compressive reconstruction;\\ Regularization parameter $\lambda$;\\ The number of NC-OFDM symbols $M_{\text{sym}}$;\\ The total number of subcarriers $N_{\text{c}}$;\end{tabular}} \\
\textbf{Output:}               & \multicolumn{4}{l}{The velocity of target $\hat{v_0}$.} \\ 
\textbf{Part 1:}         & \multicolumn{4}{l}{} \\ 
1:        & \multicolumn{4}{l}{Initialize a matrix $\mathbf{Z}\in \mathbb{C}^{N_{\text{c}}\times M_{\text{sym}}}$;}  \\
2:      & \multicolumn{4}{l}{$\textbf{For}$ $\mathbf{Z}$ $m$ in $M_{\text{sym}}$ $\textbf{do}$}   \\
3:        & \multicolumn{4}{l}{$\hspace{1em}$ The $m$-th sequence $\mathbf{A}_m$ is assigned to $\mathbf{Z}(:,m)$;}  \\
4:        & \multicolumn{4}{l}{$\textbf{End}$ $\textbf{For}$}  \\
5:       & \multicolumn{4}{l}{Obtain $\mathbf{Z}$;}  \\
6:                             & \multicolumn{4}{l}{Initialize the processed channel information matrix $\mathbf{D'}_{\mathrm{V}} \in \mathbb{C}^{N_{\text{c}}\times M_{\text{sym}}}$;}  \\
\multirow{-2}{*}{7:}                             & \multicolumn{4}{l}{\begin{tabular}[c]{@{}l@{}}The result of $\mathbf{D}_{\mathrm{V}}$ and $\mathbf{Z}$ are performed a Hadamard product \\ is assigned to $\mathbf{D'}_{\mathrm{V}}(:,:)$;\end{tabular}}  \\
8:                             & \multicolumn{4}{l}{Obtain $\mathbf{D'}_{\mathrm{V}}$;}  \\
\textbf{Part 2:}         & \multicolumn{4}{l}{} \\ 
9:                             & \multicolumn{4}{l}{Initialize a velocity power spectrum $\mathbf{Q}\in \mathbb{C}^{1 \times M_{\text{sym}}}$ and $i=1$;}  \\
10:                            & \multicolumn{4}{l}{$\textbf{For}$ $\mathbf{D'}_{\mathrm{V}}$ $n'$ in $N_{\text{c}}$ $\textbf{do}$}   \\
11:                            & \multicolumn{4}{l}{$\hspace{1em}$ $\textbf{If}$ the $n'$-th row vector of $\mathbf{D'}_{\mathrm{V}}$ $>0$ $\textbf{do}$}   \\
\multirow{-2}{*}{12:}                             & \multicolumn{4}{l}{\begin{tabular}[c]{@{}l@{}}$\hspace{2em}$ Set the regularization parameter $\lambda$ according to \\ $\hspace{2em}$ Table \ref{tab_5} and Table \ref{tab_6};\end{tabular}}   \\
13:                            & \multicolumn{4}{l}{$\hspace{2em}$ Set the maximum number of iterations $I_{\text{max}}$;}  \\
14:                            & \multicolumn{4}{l}{$\hspace{2em}$ Set the threshold of iterations error $\Theta$;}    \\
\multirow{-2}{*}{15:}                            & \multicolumn{4}{l}{\begin{tabular}[c]{@{}l@{}}$\hspace{2em}$ The $n'$-th row vector of $\mathbf{D'}_{\mathrm{V}}$ and the selection matrix $\mathbf{G_{n'}}$ \\ $\hspace{2em}$ perform (\ref{eq27}) to obtain $\mathcal{K'}_{n'}$;\end{tabular}}    \\
16:                            & \multicolumn{4}{l}{$\hspace{2em}$ Take $\mathcal{P}_1=\mathcal{Z}\in \mathbb{R}^{M_{\text{sym}}}$, $t=1$;} \\
17:                            & \multicolumn{4}{l}{$\hspace{2em}$ $\textbf{While} \ \text{error}(k+1)-\text{error}(k)\ge \Theta$ and $k \le I_{\text{max}}$ $\textbf{do}$}   \\
18:                            & \multicolumn{4}{l}{$\hspace{3em}$ $\mathcal{Z}_k = p_{L}(\mathcal{P}_k)$;}   \\
19:                            & \multicolumn{4}{l}{$\hspace{3em}$ $t_{k+1}=\frac{1+\sqrt{1+4t_{k}^{2}}}{2}$;}   \\
20:                            & \multicolumn{4}{l}{$\hspace{3em}$ $\mathcal{P}_{k+1}=\mathcal{Z}_k+\left(\frac{t_k-1}{t_{k+1}}\right)(\mathcal{Z}_k-\mathcal{Z}_{k-1})$;}   \\
21:                            & \multicolumn{4}{l}{$\hspace{3em}$ $\text{error}(k+1)=\|\mathbf{F}_{n'}\mathcal{Z}_k-\mathcal{K'}_{n'}\|_2$;}    \\
22:                            & \multicolumn{4}{l}{$\hspace{2em}$ $\textbf{End While}$}  \\
23:                            & \multicolumn{4}{l}{$\hspace{2em}$ Obtain the reconstruction data $\mathcal{Z}$;}  \\
24:                            & \multicolumn{4}{l}{$\hspace{2em}$ $\mathbf{Q}=\mathbf{Q}+|(\mathcal{Z})^{\text{T}}|$;}  \\
25:                            & \multicolumn{4}{l}{$\hspace{2em}$ $i=i+1$;}  \\
26:                            & \multicolumn{4}{l}{$\hspace{1em}$ $\textbf{End If}$}  \\
27:                            & \multicolumn{4}{l}{$\textbf{End For}$}  \\
28:                            & \multicolumn{4}{l}{$\mathbf{Q}=\mathbf{Q}/ i$;}  \\
\textbf{Part 3:}         & \multicolumn{4}{l}{} \\ 
29:                            & \multicolumn{4}{l}{Search for the peak value of $\mathbf{Q}$ and get the peak index value;} \\
30:                            & \multicolumn{4}{l}{Substitute the index value into $(\ref{eq28})$;}  \\
\hline
\end{tabular}}
\end{table}

When the input $\mathbf{y}_m$ or $\mathbf{k}_{n'}$ is a vector without zero elements, such as the first row vector under scenario 1, there is no deterioration of sidelobes due to the zero elements. However, using the FISTA can achieve the effect of noise reduction, and the noise reduction can also improve PSLR \cite{9605181}.

\section{The Selection of $\lambda$ and Sensing Performance Analysis} \label{se4}
In this section, the KCV in machine learning is applied to find the optimal value of $\lambda$, which results in a faster and more accurate reconstruction capability for JCMSA. Then, the impact of JCMSA on sensing performance, including SNR gain, resolution, and computational complexity, is analyzed.

\subsection{The Selection of $\lambda$ based on Machine Learning}
FISTA is a fast algorithm used to solve linear inverse problems and can also be applied to CS reconstruction problems. For the mathematical model of the solution of (\ref{eq4}), it can be transformed into an equivalent mathematical model \cite{kutyniok2013theory}
\begin{equation}\label{eq29}
		\min \limits_{\theta} \frac{1}{2}\|\mathbf{A}\theta-\mathbf{y}\|_2^2+\lambda\|\theta\|_1,
\end{equation}
where $\| \cdot \|_2$ represents the $\ell_2$-norm, and $\lambda$ is the regularization parameter \cite{beck2009fast}. This is a least square optimization problem with an $ \ell_1 $ regularization term. The FISTA is used for solving such problems are shown in Table \ref{tab_3}.
\begin{table}[!ht]
	\centering
	\caption{FISTA with a constant stepsize \cite{beck2009fast}.}
	\label{tab_3}
    \renewcommand{\arraystretch}{1.5}
    \resizebox{.45\textwidth}{!}{
     \begin{tabular}{|l|}
			\hline
			\textbf{Input:} $L = L(f) $ : A Lipschitz constant of $\bigtriangledown f$.\qquad \qquad \qquad	\\ \textbf{Step 0.} Take $\mathbf{y}_1 = \mathbf{x}_0 \in \mathbb{R}^n$, $t_1 = 1$.\qquad \qquad \qquad	\\
		    \textbf{Step k.} $(k \ge 1)$ Compute \qquad \qquad \qquad\\
		    \qquad \qquad \qquad  $\mathrm{x}_{k}=p_L(\mathbf{y}_{k})$,\\
		    \qquad \qquad \qquad $ t_{k+1}=\frac{1+\sqrt{1+4{t_{k}}^{2}}}{2}$,\\
		   \qquad \qquad \qquad   $\mathbf{y}_{k+1}=\mathbf{x}_{k}+(\frac{t_{k}-1}{t_{k+1}})(\mathbf{x}_{k}-\mathbf{x}_{k-1})$,\\
           where \\
            $p_L(\mathbf{y})=\underset{\mathbf{x}}{\operatorname*{argmin}}\left\{g(\mathbf{x})+\frac{L}{2}\left\Vert\mathbf{x}-\left(\mathbf{y}-\frac{1}{L}\nabla f(\mathbf{y})\right)\right\Vert^2\right\}$, \\    $g(\mathbf{x})=\lambda \|\mathbf{x}\|_1 $ ,\\
		   \hline
		\end{tabular}  }  
\end{table}

Since the $\lambda$ effects solution stability and approximation ability, choosing the appropriate regularisation parameter is beneficial to obtain the optimal reconstruction. The selection of the optimal $\lambda$  is achieved by using the KCV technique in machine learning applications \cite{han2013compressive}. 

In KCV, the dataset is divided into $\mathbf{K}$ folds, a model is learned using $\mathbf{K}-1$ folds, and an error value is calculated by testing the model in the remaining fold. Finally, the KCV estimation of the error is the average value of the errors committed in each fold \cite{rodriguez2009sensitivity}. In this paper, the training set is not used for training as the mathematical model is already given, e.g., Eq. (\ref{eq29}). We reconstruct the test set using the FISTA in each iteration and use the reconstruction error and the speed of convergence as scores to select the optimal $\lambda$.
 \begin{itemize}
	\item[$\bullet$] \textbf{Dataset selection}:
 For the input dataset, we use the processed channel information matrix as the observation dataset, e.g., Eq. (\ref{eq13}). The sparse power spectrum is also used as the ideal reconstruction dataset and the sensing matrix is used as the sensing matrix dataset, e.g., $\mathbf{J}_m$ or $\mathbf{F}_{n'}$.
\end{itemize}
 \begin{itemize}
	\item[$\bullet$] \textbf{Trade-offs for rewards}:
We use reconstruction error and convergence speed as a bonus (or score), both of which are $90$ \% and $10$ \%, respectively. The goal is to find the optimal $\lambda$ for different SNRs within an interval of $\lambda$ using on-grid searches of different step sizes.
\end{itemize}

\begin{figure}[!ht]
	\centering
	\includegraphics[width=0.45\textwidth]{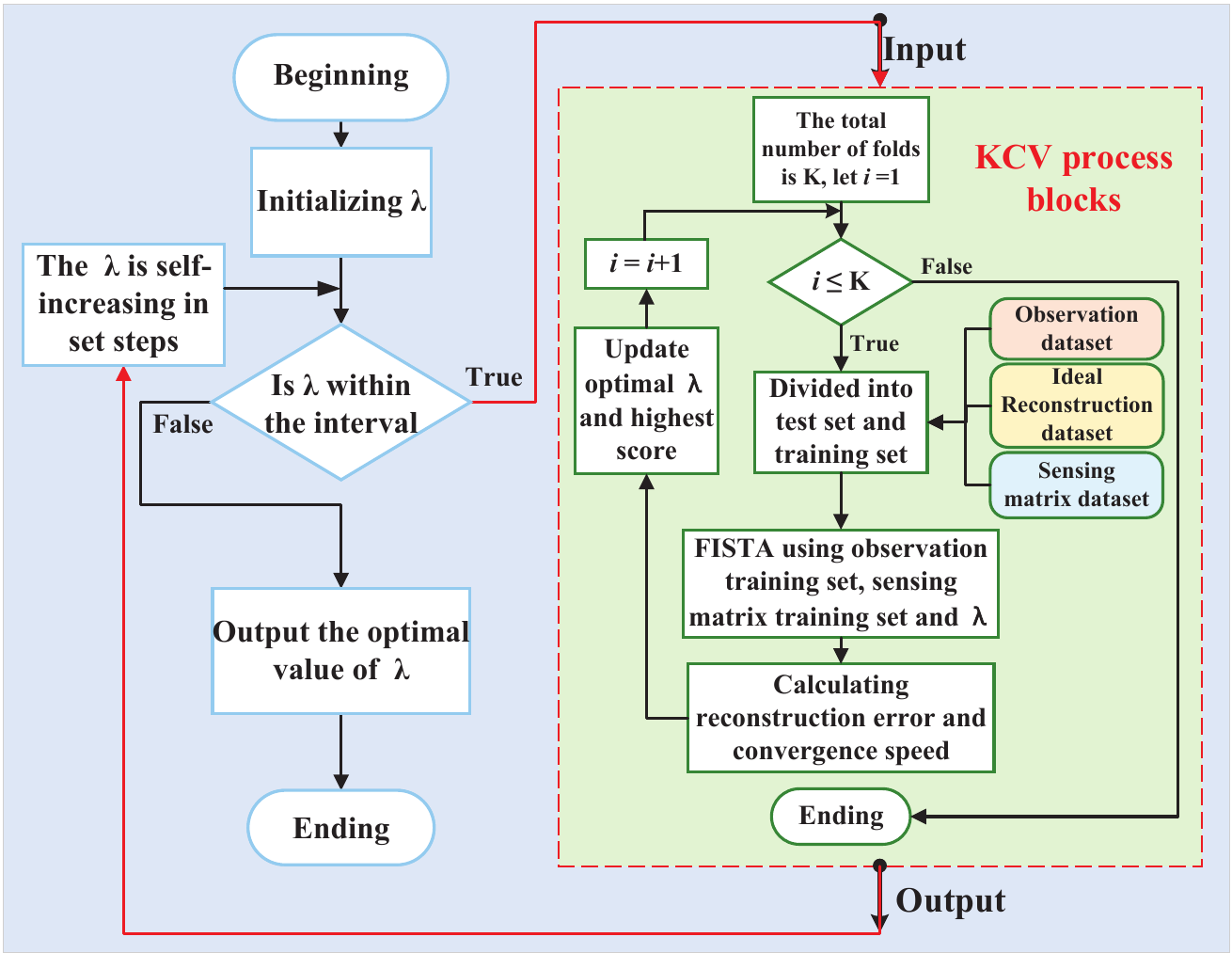}
	\caption{The flow chart for the selection of $\lambda$.}
	\label{KCV process}
\end{figure}

A detailed flow chart for the selection of $\lambda$ is shown in Fig. \ref{KCV process}.

\subsection{Sensing Performance Analysis} \label{4sce2}
\subsubsection{SNR performance} \label{4sec2.1}
An important performance metric is the SNR of radar estimation~\cite{WeiRS}. We take the channel information matrix in scenario 1 as the scenario of SNR performance analysis, and the channel information matrix in other scenarios can be similarly analyzed using the following procedures.

The SNR gain with JCMSA processing has three contributions: the gain generated by DFT or IDFT processing, the gain generated by FISTA processing, and the gain generated by non-coherent accumulation. Given that the gain from FISTA processing is related to the regularisation parameter and the Lipschitz constant \cite{beck2009fast}, which is adjusted according to the optimization model and the practical SNR situation. We assume that the gain produced by FISTA processing is $\varpi \ (\varpi \ge 0)$.

The assumptions for performance analysis are as follows.
\begin{enumerate}
    \item SNR is defined as (\ref{29.1}), where $P_\text{s}(x)$ and $P_\text{noise}$ denote the power of signal and noise, respectively. $\mathbb{N}$ is the number of sampling points of $x$ in $P_\text{s}(x)$.
    \item In scenario 1, a total of 
    $\mathcal{N}_m \ (\mathcal{N}_m<N_\text{c})$ subcarriers are available. Consider $\hat{\textbf{D}}_\mathrm{R1}$ as the matrix of unprocessed channel information under scenario 1, and the element $K_{m,n'}$ in $\hat{\textbf{D}}_\mathrm{R1}$ is written as (\ref{29.2}), where $ K'_{m,n'}$ and $\varepsilon_{m,n'}$ are the signal  and noise component in $K_{m,n'}$, respectively. 
    \item Suppose that the modulus of $K'_{m,n'}$ in each element of matrix $\hat{\textbf{D}}_\mathrm{R1}$ is $h$, while the variance of noise component $\varepsilon_{m,n'}$ is $h^2 \sigma ^2$.
\end{enumerate}
\begin{equation}\label{29.1}
    \begin{cases}
           \text{SNR}= \dfrac{P_\text{s}(x)}{P_\text{noise}}, \\[7pt]
           P_\text{s}(x) =\dfrac{\sum_{\mu=0}^{\mathbb{N}-1}\left|x(\mu)\right|^2}{\mathbb{N}}, \\[7pt]
           x~ $is the discrete signal sequence.$
    \end{cases}   
\end{equation}
\begin{equation}\label{29.2}
    K_{m,n'}=
    \begin{cases}
        \varepsilon_{m,n'} ,  K_{m,n'} $\ in licensed point$, \\
        K'_{m,n'} + \varepsilon_{m,n'},  K_{m,n'} $\ in unlicensed point $.
    \end{cases}
\end{equation}

{\setlength{\parindent}{0pt} \begin{theorem}\label{theorem1}
 When FISTA finds the optimal parameter $\lambda$, the SNR gain with FISTA is $\varpi > 0 $, and hence we have
 \begin{equation} \label{eq32.2}
        \begin{cases}
            \underbrace{\sqrt{M_{\text{sym}}}\left(\frac{\mathcal{N}_m}{\sigma^2}+ \varpi \right)}_{\text{ JCMSA}} > \underbrace{\frac{\sqrt{M_{\text{sym}}}\mathcal{N}_m}{\sigma^2}}_{\text{ \cite{wei2022robust}}} > \underbrace{\frac{\mathcal{N}_m^2}{N_\text{c} \sigma^2}}_{\text{2D FFT}} ,\ (\mathrm{range}) \\
            \underbrace{\sqrt{\mathcal{N}_m}\left(\frac{M_{\text{sym}}}{\sigma^2}+\varpi \right)}_{\text{ JCMSA}} > \underbrace{\frac{\sqrt{\mathcal{N}_m}M_{\text{sym}}}{\sigma^2}}_{\text{ \cite{wei2022robust}}} >
            \underbrace{\frac{M_{\text{sym}}}{\sigma^2}}_{\text{2D FFT}}. \ (\mathrm{velocity})
        \end{cases}
\end{equation}
\end{theorem}}

\begin{proof}    
In the following, we first derive the SNR gains for range and velocity with JCMSA processing, and then give the SNR gains with conventional 2D FFT processing and the SNR gains with improved algorithm processing in \cite{wei2022robust}. The results show that the JCMSA proposed in this paper has better SNR performance gain.

The SNR gain of range with JCMSA processing is derived as follows.
\begin{itemize}
    \item Step 1: 
    After the \textbf{Part 1} in \hyperref[tab1]{Algorithm 1}, each column vector of matrix $\hat{\textbf{D}}_\mathrm{R1}$ contains $\mathcal{N}_m$ signal elements and $\mathcal{N}_m$ noise elements.
    \item Step 2:
    The \textbf{Part 2} in \hyperref[tab1]{Algorithm 1} contains three operations: IDFT, FISTA, and non-coherent accumulation. After the IDFT on each column vector, the SNR of each column vector is $\mathcal{N}_m/\sigma^2$. Then using FISTA, the SNR is improved to $(\mathcal{N}_m/\sigma ^2)+\varpi$. 
    \item Step 3: 
    It is not possible to directly compute an SNR after non-coherent accumulation from \cite{richards2010noncoherent}. Therefore, in this paper, the non-coherent accumulation gain is assumed to be $\sqrt{E}$, where $E$ is the number of symbols undergoing non-coherent accumulation. In scenario 1, there are $M_{\text{sym}}$ symbols for non-coherent accumulation. Thus, after the non-coherent accumulation, the SNR of output signal is $\sqrt{M_{\text{sym}}}\left[(\mathcal{N}_m/\sigma ^2)+\varpi\right]$.
\end{itemize}

The SNR gain of velocity with JCMSA processing is derived as follows.
\begin{itemize}
    \item Step 1:
    After the \textbf{Part 1} in \hyperref[tab2]{Algorithm 2}, the noise elements of the unavailable subcarrier positions in $\hat{\textbf{D}}_\mathrm{R1}$ are 
    zeroed out.
   \item Step 2:
   The \textbf{Part 2} in \hyperref[tab2]{Algorithm 2} contains three operations: DFT, FISTA, and non-coherent accumulation. After the DFT on each row vector, the SNR of each row vector is $M_{\text{sym}}/\sigma^2$. Then after FISTA, the SNR is improved to $(M_{\text{sym}}/\sigma ^2)+\varpi$.
   \item Step 3: 
   Unlike the non-coherent accumulation of range, the number of accumulations is the number of available subcarriers, i.e., $\mathcal{N}_m$. Thus, after the non-coherent accumulation, the SNR of output signal is $\sqrt{\mathcal{N}_m}\left[(M_{\text{sym}}/\sigma ^2)+\varpi\right]$.
\end{itemize}

For the conventional 2D FFT algorithm, IDFT is performed on a column of $\hat{\textbf{D}}_\mathrm{R1}$ 
to obtain the range power spectrum, and DFT is performed on a row to obtain the velocity power spectrum. Thus, the SNR for range and velocity of the output signal after the conventional 2D FFT algorithm are $\mathcal{N}_m^2/(N_\text{c} \sigma^2)$ and $M_{\text{sym}}/\sigma^2$,  respectively.

For the improved algorithm in \cite{wei2022robust}, only the FISTA gain is missing compared to JCMSA. Hence, the total SNR gain results are just missing $\varpi$.
\end{proof}
{\setlength{\parindent}{0pt}
\begin{theorem}\label{theorem2}
    When the input signal-to-noise ratio is lower than -15 dB, it is difficult to find the optimal FISTA parameter $\lambda$, i.e. $\varpi \approx 0$. Thus, we have 
    \begin{equation} \label{eq33.2}
        \begin{cases}
            \underbrace{\sqrt{M_{\text{sym}}}\left(\frac{\mathcal{N}_m}{\sigma^2}  \right)}_{\text{ JCMSA}} \overset{\Delta}{\operatorname*{=}}\underbrace{\frac{\sqrt{M_{\text{sym}}}\mathcal{N}_m}{\sigma^2}}_{\text{ \cite{wei2022robust}}} > \underbrace{\frac{\mathcal{N}_m^2}{N_\text{c} \sigma^2}}_{\text{2D FFT}}, \ (\mathrm{range}) \\
            \underbrace{\sqrt{\mathcal{N}_m}\left(\frac{M_{\text{sym}}}{\sigma^2} \right)}_{\text{ JCMSA}} \overset{\Delta}{\operatorname*{=}}\underbrace{\frac{\sqrt{\mathcal{N}_m}M_{\text{sym}}}{\sigma^2}}_{\text{ \cite{wei2022robust}}} >
            \underbrace{\frac{M_{\text{sym}}}{\sigma^2}}_{\text{2D FFT}}. \ (\mathrm{velocity})
        \end{cases}
\end{equation}
\end{theorem}}

\begin{proof}
    Observing the (\ref{eq32.2}) in \hyperref[theorem1]{\text{Theorem 1}}, we notice that the only difference between the SNR gains of JCMSA and improved algorithm in \cite{wei2022robust} is $\varpi$. Therefore, When $\varpi \approx 0$, the JCMSA and improved algorithm in \cite{wei2022robust} have similar SNR gains, both of which perform better than the conventional 2D FFT algorithm.
\end{proof}

\subsubsection{Resolution performance}
\begin{figure}[htbp]
	\centering
	\subfigure[Continuous occupancy of $N_\text{c}/2$ subcarriers within a sensing frame.] {\label{fig6.a}\includegraphics[width=.24\textwidth]{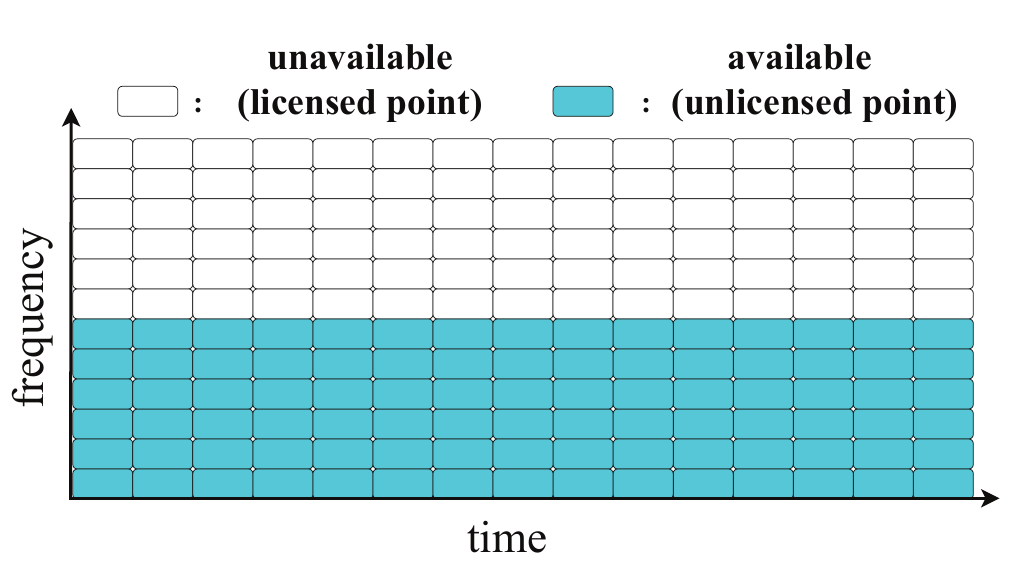}}
	\subfigure[Continuous occupancy of $M_{\text{sym}}/2$ NC-OFDM symbols within a sensing frame.] {\label{fig6.b}\includegraphics[width=.24\textwidth]{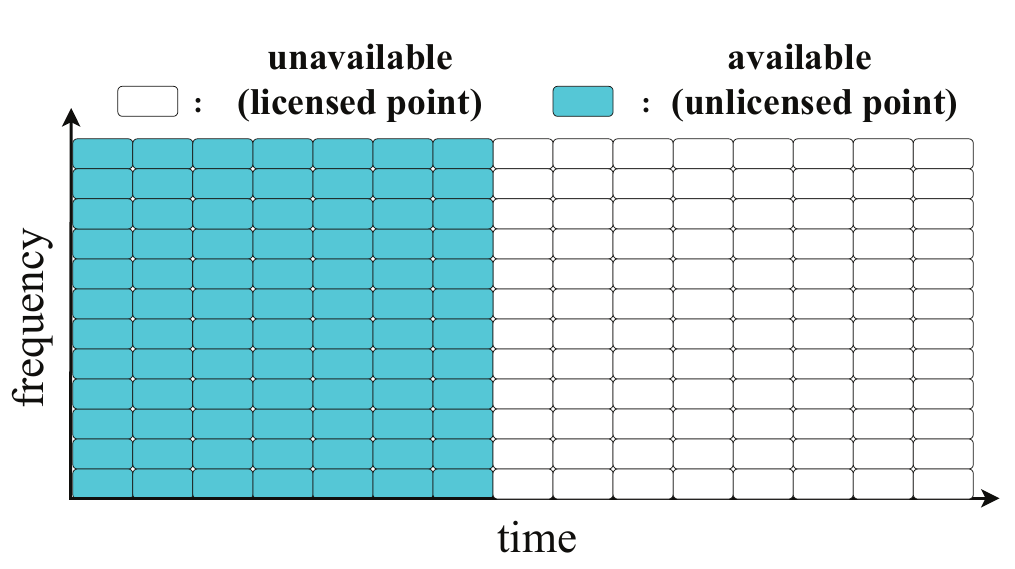}}
	\caption{Dynamic changes in spectrum occupancy.}
	\label{fig.6}
\end{figure}

As depicted in Fig. \ref{fig.6}, two extreme cases of spectrum occupancy have been selected to facilitate the analysis of resolution.

For conventional 2D FFT algorithm, the resolution of the range is not affected by the waveform used or the specific parameterization of the NC-OFDM system~\cite{WeiIN}. It is solely determined by the bandwidth occupied by the transmit signal\cite{sturm2009novel}. Therefore, When the spectrum occupancy of the ISAC system is distributed as shown in Fig. \ref{fig6.a}, the range resolution of the conventional 2D FFT algorithm is
\begin{equation}\label{30}
	\Delta R_{\text{2dfft}}=\frac{c}{\Delta f\times N_{\text{c}}} \ (\text{2D FFT algorithm}).
\end{equation}
Given that the velocity resolution is related to the total duration of the sensing symbols. Therefore, when the spectrum occupancy of the ISAC system is distributed as  shown in Fig. \ref{fig6.b}, the velocity resolution of the conventional 2D FFT algorithm is
\begin{equation}\label{31}
     \Delta v_{\text{2dfft}}=\frac{c}{T_{\text{sym}}f_\text{c} M_{\text{sym}}} \ (\text{2D FFT algorithm}).
\end{equation}

In the JCMSA-enabled ISAC system working on unlicensed spectrum bands, the sensing performance under full bandwidth or full sensing symbols can be obtained, thereby maintaining the maximum resolution, e.g., Eq. (\ref{eq31.1}).
\begin{align} \label{eq31.1}
    \text{JCMSA}
    \begin{cases}
        \Delta R_{\text{jcmsa}}=\dfrac{c}{2\Delta f N_{\text{c}}},\\
        \Delta v_{\text{jcmsa}}=\dfrac{c}{2T_{\text{sym}}f_\text{c} M_{\text{sym}}}.
    \end{cases}
\end{align}

\subsubsection{Computational complexity}
As one of the performance metrics, the computational complexity is equally treated as a reflection of the sensing performance.
The computational complexity of the three algorithms will be given and analyzed as follows.
\begin{itemize}
    \item \textbf{2D FFT method:} The computational complexity of range and velocity estimation is $\mathcal{O}\left(N_\text{c}^2\right)$ and $\mathcal{O}\left(M_\text{sym}^2\right)$, respectively.
    \item \textbf{The method in \cite{wei2022robust}:} The computational complexity of range and velocity estimation is $\mathcal{O}\left(N_\text{c}^2M_\text{sym}\right)$ and $\mathcal{O}\left(M_\text{sym}^2N_\text{c}\right)$, respectively. Compared to 2D FFT method, this method increases the computational complexity of coherent accumulation operation.
    \item \textbf{JCMSA:} Different from the previous two algorithms, the JCMSA needs to analyze the computational complexity of FISTA. For an input matrix of $\mathcal{N} \times \mathcal{N}$, the computational complexity of FISTA is $\mathcal{O}\left(\mathcal{N}^2N_\text{iter}\right)$, which includes gradient descent operation, soft threshold calculation and update iteration operation, where $N_\text{iter}$ represents the number of iterations. Therefore, the computational complexity of range and velocity estimation for JCMSA is $\mathcal{O}\left(N_\text{c}^2M_\text{sym}N_\text{iter}\right)$ and $\mathcal{O}\left(M_\text{sym}^2N_\text{c}N_\text{iter}\right)$, respectively. 
\end{itemize}

The above analysis shows that JCMSA has the highest computational complexity. For algorithm design, high sensing performance often requires sacrificing other aspects of performance. Therefore, the method proposed in this paper is to sacrifice the computational complexity of the algorithm for better sensing performance.

\begin{table*}[!htbp]
	\caption{Simulation parameters \cite{wei2022robust,wei2023carrier,WeiIN,3gpp2018nr}}
	\label{tab_4}
	\renewcommand{\arraystretch}{1.2} 
	\begin{center}
    \resizebox{.6\textwidth}{!}{
		\begin{tabular}{|m{0.15\textwidth}<{\centering}| m{0.4\textwidth}<{\centering}| m{0.15\textwidth}<{\centering}|}
			\hline
			\textbf{Symbol}	& \textbf{Parameter} &\textbf{ Value } \\
			\hline
			$N_\text{c}$	& Number of subcarriers & $512$  \\
			\hline
			$\mathcal{N}_m$	& Number of occupied subcarriers & $256$  \\
			\hline
			$M_{\text{sym}}$	& Number of symbols  & $14$ \\
			\hline
			$f_\text{c}$	& Carrier frequency & 24 GHz \\
			\hline
			$\Delta f$	& Subcarrier spacing  & 15 KHz  \\
			\hline
			$T_{\text{ofdm}}$	& Elementary NC-OFDM symbol duration  & 66.67 $\mu$s  \\
			\hline
			$T_{\text{cp}}$	& Cyclic prefix length & 16.67 $\mu$s  \\
			\hline
			$T_{\text{sym}}$	& Total duration of NC-OFDM symbol & 83.34 $\mu$s  \\
			\hline
			$R$	& Range of the target & 117 m  \\
			\hline
			$v$	& Velocity of the target & 13 m/s \\
                \hline
               $\text{K}$ & Number of folds in KCV & 14 \\
			\hline
		\end{tabular}}
	\end{center}
\end{table*}

\section{Simulation Results and Analysis}\label{se5}

In this section, NC-OFDM ISAC signal is applied to sense a surrounding target 
with a relative velocity of $13$ m/s and a relative range of $117$ m. 
The simulation parameters are shown in Table \ref{tab_4}.
In order to show the performance improvement of the JCMSA proposed in this paper, 
we evaluate the performance of the parameter estimation of target 
in terms of both power spectrum and root mean square error (RMSE), respectively, 
and compare the JCMSA with the improved algorithm in \cite{wei2022robust} and the conventional 
2D FFT algorithm.

In the simulation, the optimal regularization parameter $\lambda$ for 
different SNRs is obtained by $14$-fold KCV. 
Table \ref{tab_5} and Table \ref{tab_6} show the optimal parameters of range and velocity 
estimation for different spectrum occupancy scenarios, respectively. 
Simulation results for the power spectra and RMSE are revealed as follows.

\begin{figure*}
\centering
\begin{minipage}[t]{0.49\linewidth}	
\centering
\captionsetup{type=table}	
 \caption{Optimal $\lambda$ for target estimation under scenario 1. 
 The step size of $\lambda$ for range estimation is 100 and the interval is [1, 10000]. 
 The step size of $\lambda$ for velocity estimation is 0.02 and the interval is [1, 5].}
 \label{tab_5}
 \renewcommand{\arraystretch}{2} 
 \resizebox{\textwidth}{!}
       {
		\begin{tabular}{|c|c|c|c|c|c|c|c|c|c|c|c|}
			\hline
			\multicolumn{12}{|c|}{Range estimation under scenario 1}      \\ \hline
			SNR (dB)       & 0 & 1 & 2 & 3 & 4 & 5 & 6 & 7 & 8 & 9 & 10 \\ \hline
			optimal $\lambda$ &  5401 &  5601 &  5001 &  4601 & 5401  &  5201 & 5001  & 5601  & 5601  & 5601  & 5201   \\ \hline
			\multicolumn{12}{|c|}{Velocity estimation under scenario 1}   \\ \hline
			SNR (dB)       & 0 & 1 & 2 & 3 & 4 & 5 & 6 & 7 & 8 & 9 & 10 \\ \hline
			optimal $\lambda$ &  2.16 &  1.56 & 1.54  & 1.08  &  1.2 & 1.12  &  0.74 & 0.68  & 1.16  & 1.7  & 1.5   \\ \hline
		\end{tabular}%
	}
\end{minipage}
\begin{minipage}[t]{.49\linewidth}
\centering
\captionsetup{type=table}
 \caption{Optimal $\lambda$ for target under scenario 2. 
 The step size of $\lambda$ for range is 100 and the interval is [1, 10000]. 
 The step size of $\lambda$ for velocity is 0.02 and the interval is [1, 5].}
 \label{tab_6}
 \vspace{0.46cm} 
 \renewcommand{\arraystretch}{2} 
 \resizebox{\textwidth}{!}
       {
		\begin{tabular}{|c|c|c|c|c|c|c|c|c|c|c|c|}
			\hline
			\multicolumn{12}{|c|}{Range estimation under scenario 2}                                      \\ \hline
			SNR (dB)       & 0    & 1    & 2    & 3    & 4    & 5    & 6    & 7    & 8    & 9    & 10   \\ \hline
			optimal $\lambda$ &  2501   & 3501  & 4501 & 3001 & 5001  &  4001  & 4801  &   5101  & 5201  &   5601   & 5101 \\ \hline
			\multicolumn{12}{|c|}{Velocity estimation under scenario 2}                                   \\ \hline
			SNR (dB)       & 0    & 1    & 2    & 3    & 4    & 5    & 6    & 7    & 8    & 9    & 10   \\ \hline
			optimal $\lambda$ & 1.32 & 1.70 & 1.44 & 1.28 & 0.92 & 1.20 & 1.10 & 1.46 & 1.70 & 1.56 & 1.54 \\ \hline
		\end{tabular}%
	}
\end{minipage} 
	{\noindent} \rule[-10pt]{18cm}{0.05em}
\end{figure*}

\subsection{Power Spectra}
In this subsection, we use the parameters in Table \ref{tab_4} in simulation, 
revealing the range and velocity power spectra for two scenarios.

\begin{figure}[!htbp]
	\centering
	\subfigure[Range estimate at SNR of 10 dB.] {\label{fig7.a}\includegraphics[width=.35\textwidth]{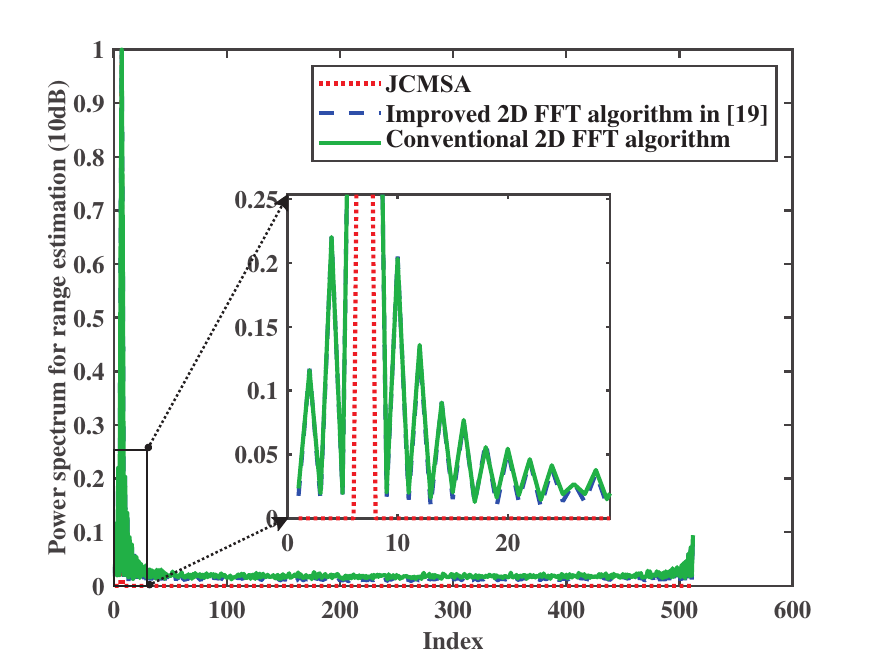}}
	\subfigure[Velocity estimate at SNR of 10 dB.] {\label{fig7.b}\includegraphics[width=.35\textwidth]{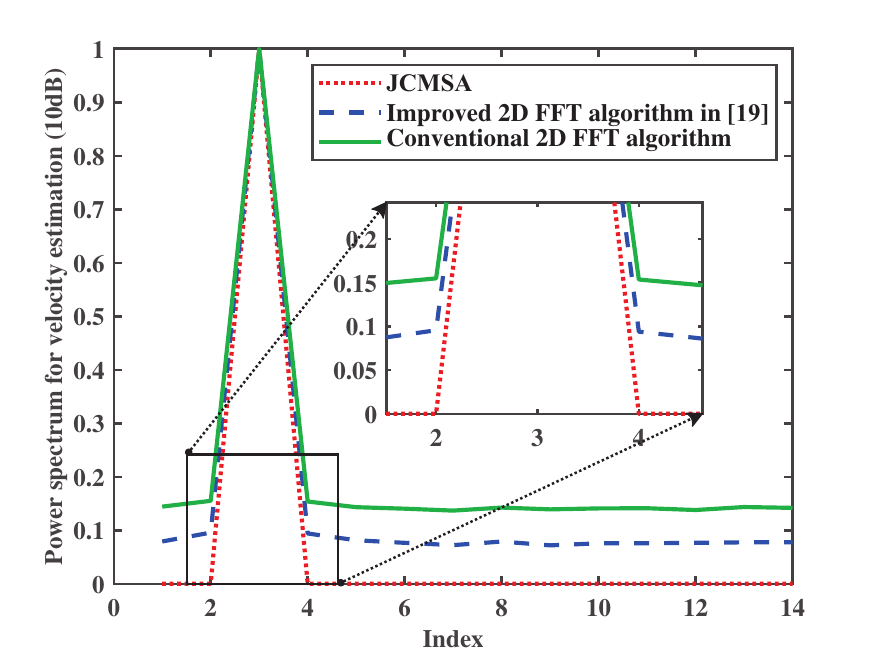}}
	\caption{Power spectrum under scenario 1.}
	\label{fig.7}
\end{figure}

\begin{figure}[!htbp]
	\centering
	\subfigure[Range estimate at SNR of 10 dB.] {\label{fig8.a}\includegraphics[width=.35\textwidth]{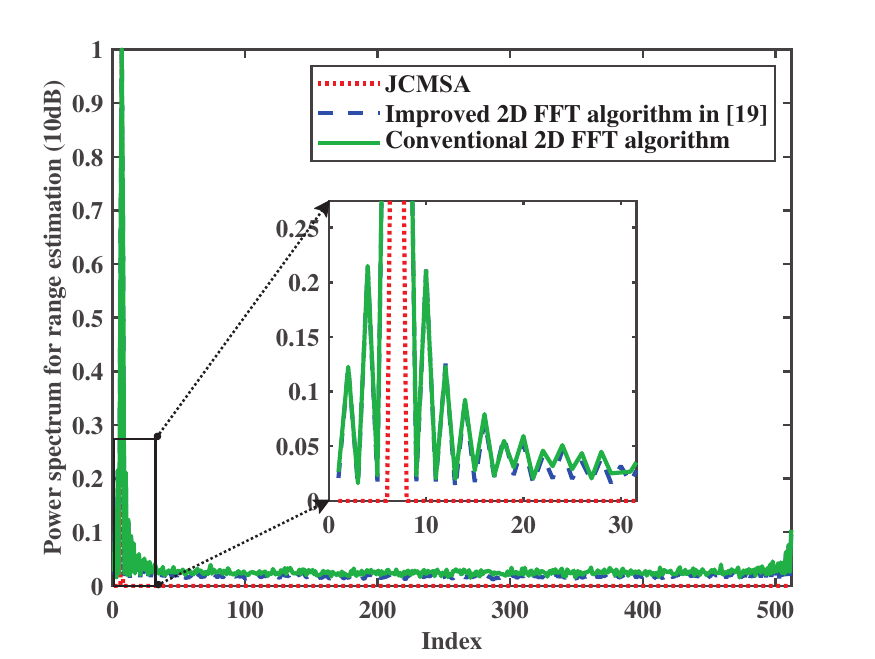}}
	\subfigure[Velocity estimate at SNR of 10 dB.] {\label{fig8.b}\includegraphics[width=.35\textwidth]{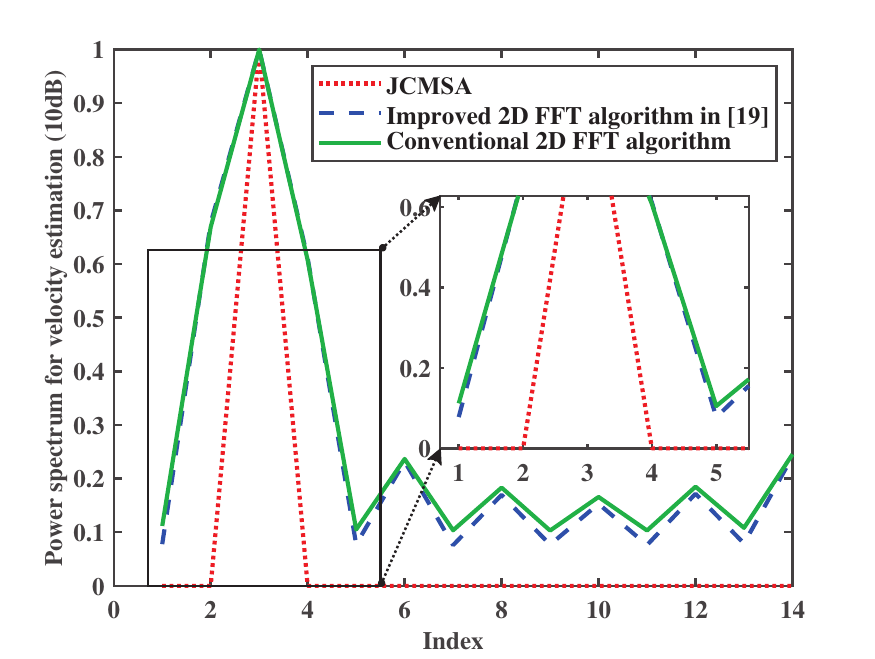}}
	\caption{Power spectrum under scenario 2.}
	\label{fig.8}
\end{figure}

Fig. \ref{fig.7} and Fig. \ref{fig.8} show the results of the power spectra 
for scenario 1 and scenario 2 respectively. 
The simulation results show that the peak index value of the range power spectrum is $\text{ind}_r =7$ 
and the peak index value of the velocity power spectrum is $\text{ind}_v =3$. 
The estimated range of the target is 117.1875 m and the estimated velocity is 13.3929 m/s 
when substituted into (\ref{eq21}) and (\ref{eq28}), respectively. 
Meanwhile, it is revealed from Fig. \ref{fig7.a}, Fig. \ref{fig8.a}, and Fig. \ref{fig8.b} 
that the spectral discontinuity worsens the power spectral sidelobes, and the worst PSLR is -6.4 dB, 
and the improved 2D FFT algorithm in \cite{wei2022robust} cannot solve this problem.
while the JCMSA proposed in this paper can always maintain zero sidelobes.
For the case shown in Fig. \ref{fig7.b}, the sidelobe deterioration does not occur as 
there are no spectrum holes when velocity estimation is performed. 
However, the JCMSA proposed in this paper is able to perform a denoising effect and 
improve the PSLR to a greater level than other algorithms.

\subsection{RMSE}

RMSE is the parameter used to reveal the error between the estimated value and the real value~\cite{WeiMI}. 
With the same simulation parameters, this subsection compares the RMSE of range and 
velocity estimation under the three algorithms above. 
In this simulation, 
there exist theoretical upper and lower bounds for the RMSE of range and velocity  
due to the influence of spectrum resources and algorithms. 
According to the simulation parameters, (\ref{eq21}), and (\ref{eq28}), 
the upper and lower bounds are obtained as follows.
\begin{subequations}
    \begin{align}     
        &\overbrace{\textbf{RMSE-R}}^{\rm{upper \ bound}}= \dfrac{c(N_\text{c}-1)}{2N_\text{c} \Delta f}-117= 0.9863 \times 10^4 \ \text{m} , \label{35.a}\\
        &\underbrace{\textbf{RMSE-R}}_{\rm{lower \ bound}}=
        \dfrac{c(\text{ind}_n -1)}{2N_\text{c} \Delta f}-117= 0.1875 \ \text{m} \label{35.b},
    \end{align}
\end{subequations}

\begin{subequations} 
 \begin{align}        
        &\overbrace{\textbf{RMSE-V}}^{\rm{upper \ bound}}= \dfrac{c(M_{\text{sym}}-1)}{2M_{\text{sym}}T_{\text{sym}} f_\text{c}}-13= 56.56 \ \text{m/s} ,\label{36.a}\\
        &\underbrace{\textbf{RMSE-V}}_{\rm{lower \ bound}}=
        \dfrac{c(\text{ind}_m -1)}{2M_{\text{sym}}T_{\text{sym}} f_\text{c}}-13= 0.3929 \ \text{m/s} \label{36.b}.
\end{align}   
\end{subequations}

\begin{figure}[!htbp]
	\centering
	\subfigure[RMSE for range estimation.] {\label{fig9.a}\includegraphics[width=.35\textwidth]{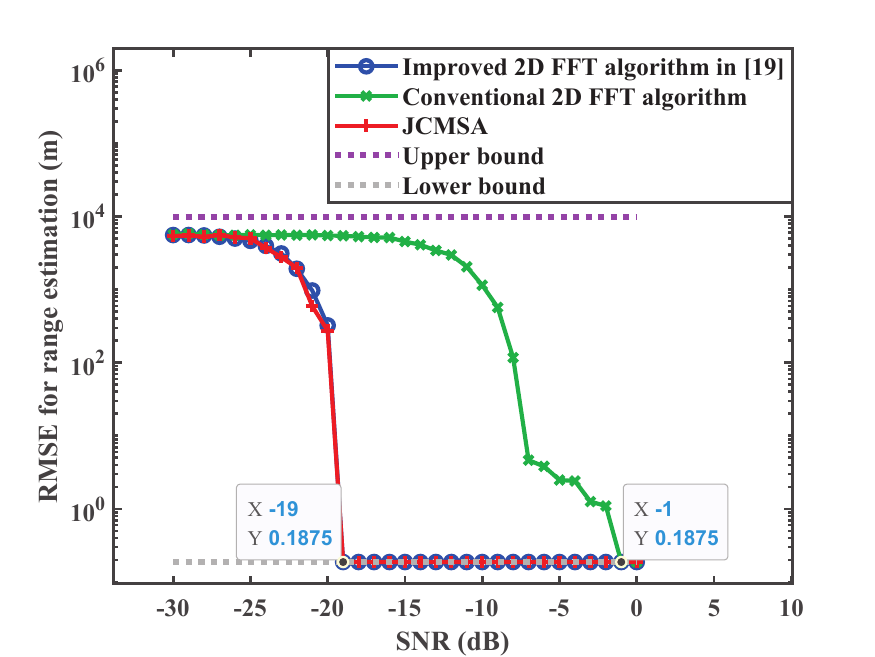}}
	\subfigure[RMSE for velocity estimation.] {\label{fig9.b}\includegraphics[width=.35\textwidth]{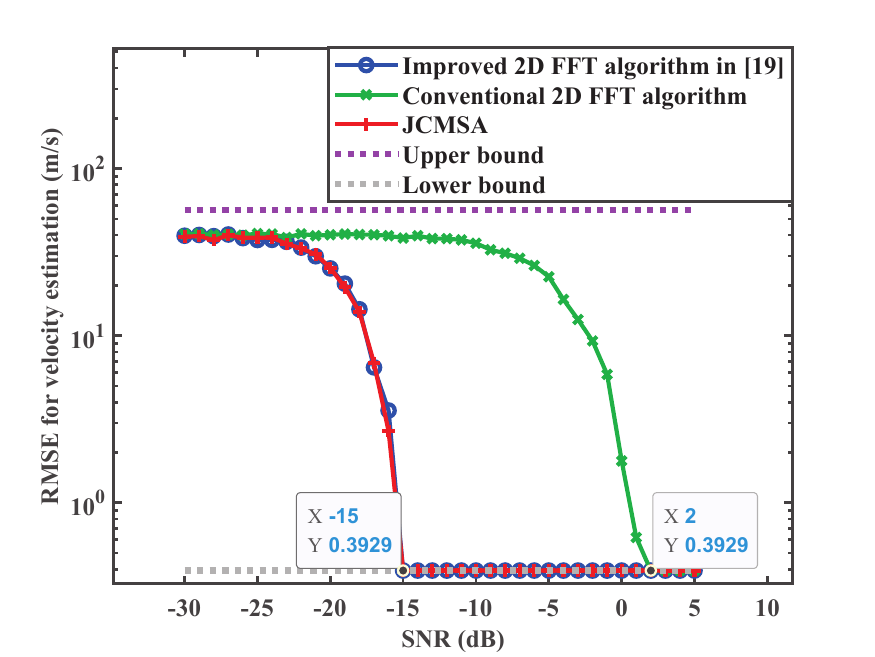}}
	\caption{RMSE under scenario 1.}
	\label{fig.9}
\end{figure}

\begin{figure}[!htbp]
	\centering
	\subfigure[RMSE for range estimation.] {\label{fig10.a}\includegraphics[width=.35\textwidth]{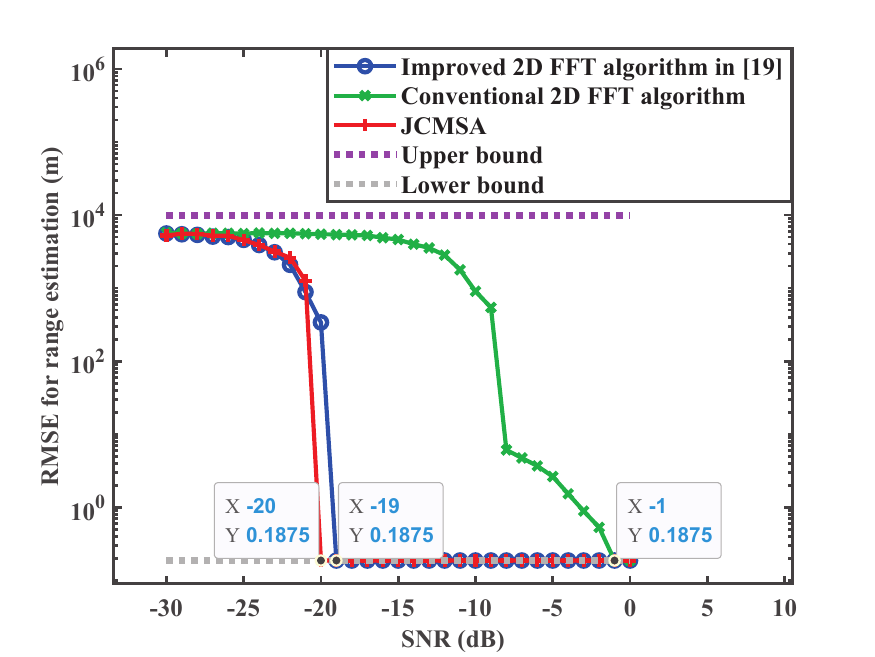}}
	\subfigure[RMSE for velocity estimation.] {\label{fig10.b}\includegraphics[width=.35\textwidth]{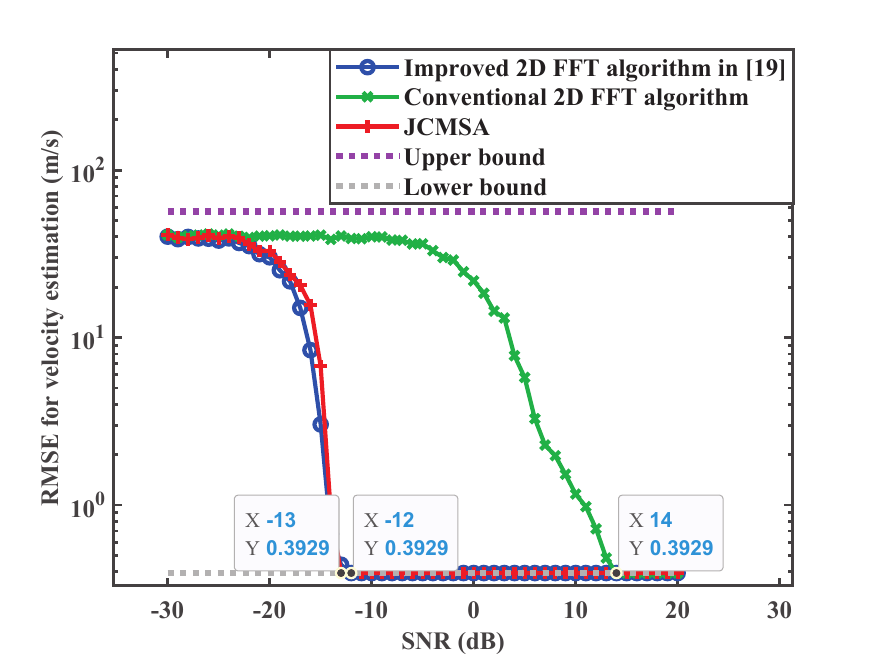}}
	\caption{RMSE under scenario 2.}
	\label{fig.10}
\end{figure}

Fig. \ref{fig.9} and Fig. \ref{fig.10} show the results of the RMSEs of range and 
velocity estimations under scenario 1 and scenario 2, respectively. 
According to the simulation results, three conclusions are drawn as follows.
\begin{enumerate}
    \item The JCMSA and the improved algorithm in \cite{wei2022robust} have similar anti-noise performance. The anti-noise performance and SNR gain are correlated. When the input SNR is low, the JCMSA and improved algorithm in \cite{wei2022robust} have similar SNR gains, which is proved in \hyperref[theorem2]{\text{Theorem 2}}. Thus, both algorithms show similar anti-noise performance when the SNR is less than $-15$ dB.
    \item The JCMSA has better anti-noise performance than conventional 2D FFT algorithms, because the SNR is larger than conventional 2D FFT algorithm with JCMSA, 
    which is proved in \hyperref[theorem1]{\text{Theorem 1}} and \hyperref[theorem2]{\text{Theorem 2}}. 
    \item When the SNR is below $-25$ dB, the estimated RMSE is mainly affected by noise, so that the average error of the RMSE hardly reaches the theoretical upper bound ((\ref{35.a}), (\ref{36.a})). When the SNR is higher than a certain value, the estimated RMSE depends on the resolution. The theoretical lower bound of the RMSE is also determined by the resolution, so that the estimated RMSE can reach the theoretical lower bound ((\ref{35.b}), (\ref{36.b})).
\end{enumerate}


Since all of them perform periodogram estimation, the RMSE will reach the lower bound when SNR is high. Therefore, the performance comparison of the three methods in the case of high SNR cannot be obtained. To this end, we break the lower bound of the periodogram by matching the obtained data with the delineated grid \cite{wei2023symbol}, thereby obtaining the performance comparison of the three methods in the case of high SNR. It should be declared that this operation was performed for all three methods, so that the comparison of the three methods is still fair.

\begin{figure}[!htbp]
	\centering
	\subfigure[RMSE for range estimation.] {\label{fig11.a}\includegraphics[width=.35\textwidth]{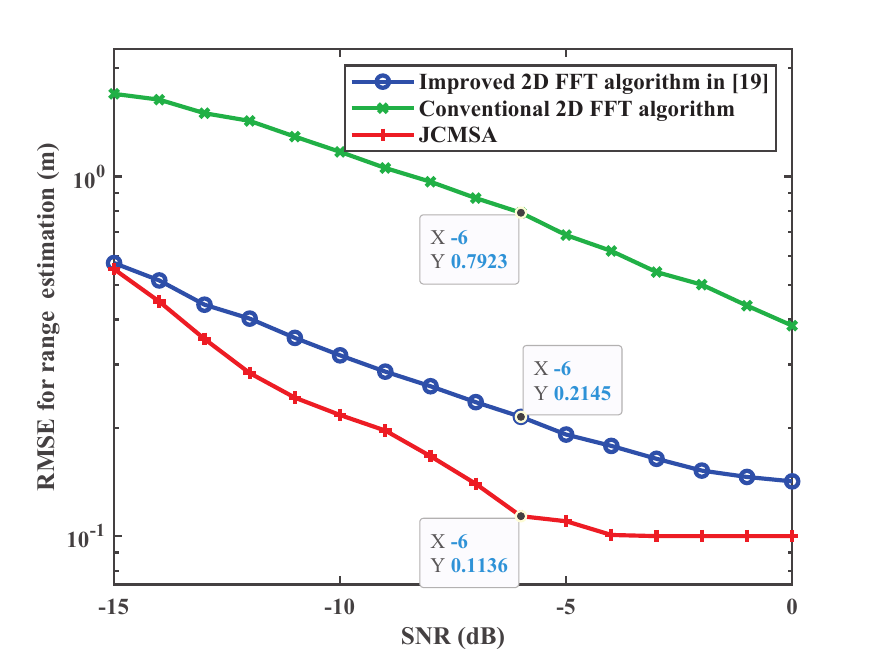}}
	\subfigure[RMSE for velocity estimation.] {\label{fig11.b}\includegraphics[width=.35\textwidth]{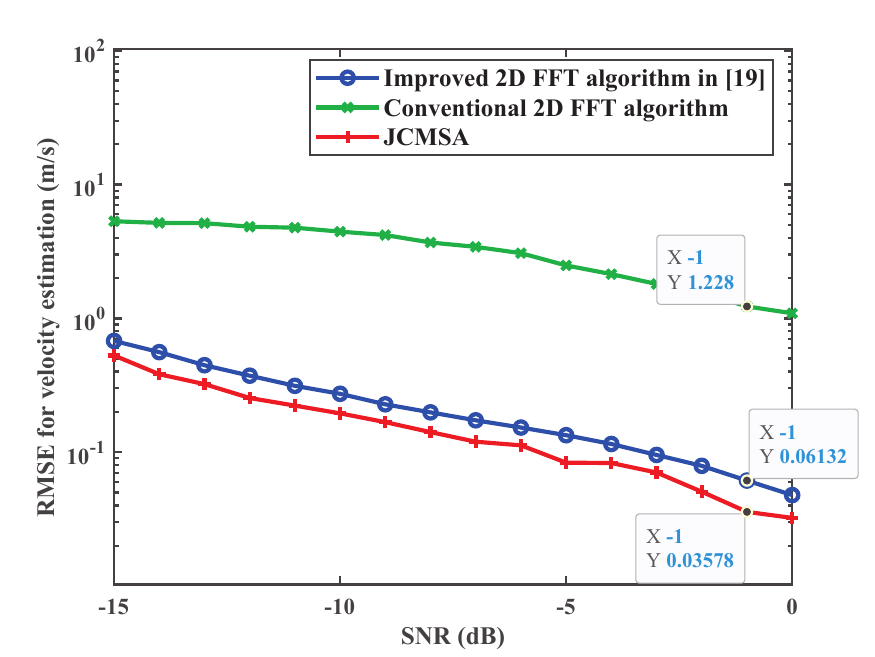}}
	\caption{RMSE under scenario 1.}
	\label{fig.11}
\end{figure}

\begin{figure}[!htbp]
	\centering
	\subfigure[RMSE for range estimation.] {\label{fig12.a}\includegraphics[width=.35\textwidth]{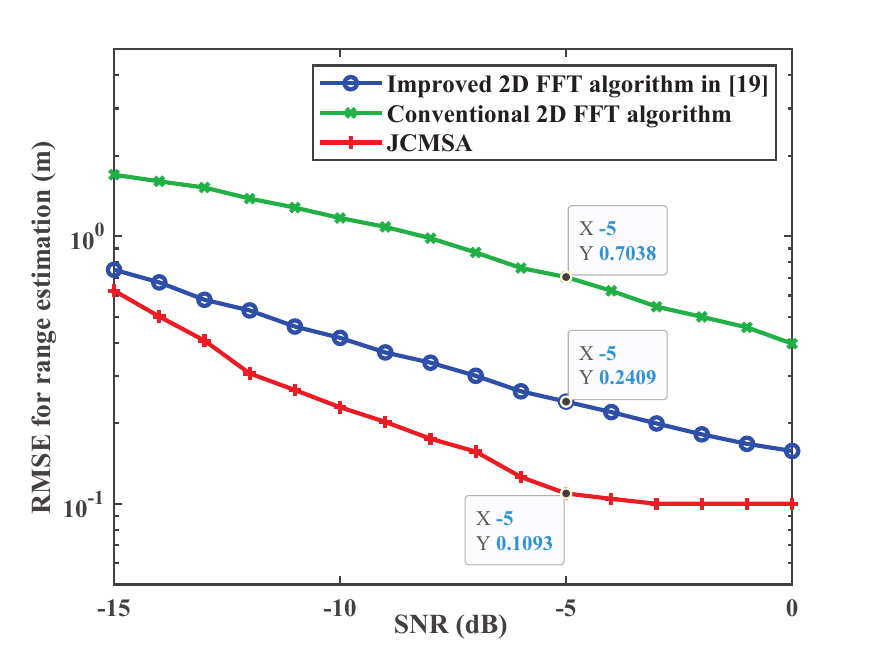}}
	\subfigure[RMSE for velocity estimation.] {\label{fig12.b}\includegraphics[width=.35\textwidth]{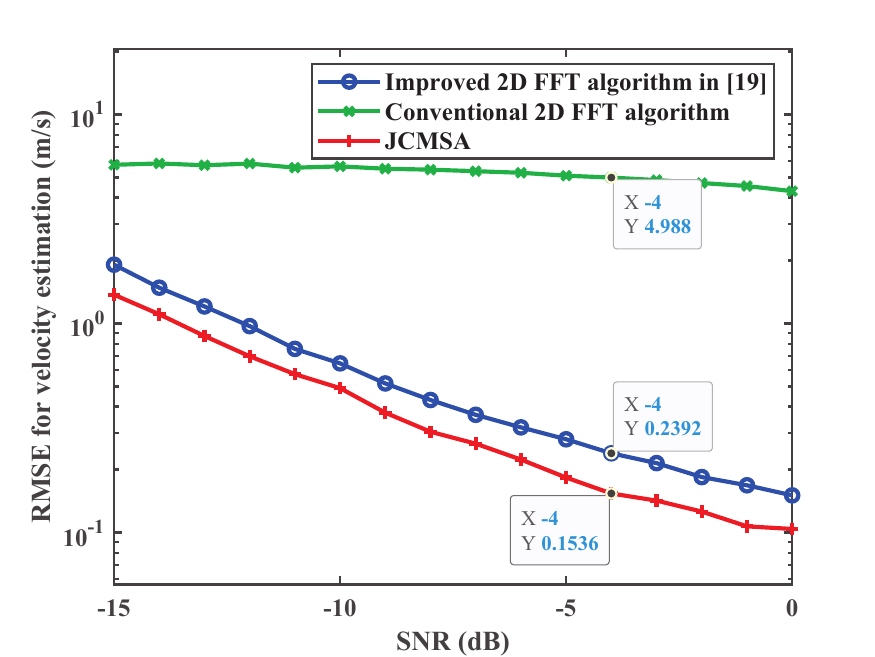}}
	\caption{RMSE under scenario 2.}
	\label{fig.12}
\end{figure}

Fig. \ref{fig.11} and Fig. \ref{fig.12} show the results of RMSEs of range and velocity estimation with SNR greater than -15 dB. As shown in Fig. \ref{fig11.a} and Fig. \ref{fig12.a}, the accuracy of range estimation is maximally improved by 54.66 \%, 84.36 \%, compared with improved 2D FFT algorithm in \cite{wei2022robust} and conventional 2D FFT algorithm, respectively. As shown in Fig. \ref{fig11.b} and Fig. \ref{fig12.b}, the accuracy of velocity estimation is maximally improved by 41.54 \%, 97.09 \%, compared with improved 2D FFT algorithm in \cite{wei2022robust} and conventional 2D FFT algorithm, respectively. Furthermore, the accuracy of range and velocity estimation with the proposed algorithm in this paper is satisfy the required for current scenarios \cite{rong20216g}. Simulation results prove that the JCMSA is superior with a high SNR, which confirms with \hyperref[theorem1]{\text{Theorem 1}}.


\subsection{Computational complexity}
We have operated the three methods on simulation equipment, where the CPU parameters of the simulation device is 13-th Gen Intel (R) Core (TM) i9-13900H. In addition, the number of iterations of FISTA is $N_\text{iter}=1\times10^3$, and the running time of the three methods under the above device is given in Table \ref{tab_7}.

Table \ref{tab_7} shows that the computational complexity of JCMSA is much higher than the other two methods, which is consistent with the theoretical analysis. The above running time will be lower in the actual equipment and will not affect the normal operation of the equipment.

\begin{table}[ht]
\caption{The computational complexity of sensing method }
\label{tab_7}
\renewcommand{\arraystretch}{1.5} 
\begin{center}
\resizebox{.45\textwidth}{!}{
\begin{tabular}{|cccc|}
\hline
\multicolumn{4}{|c|}{Computational complexity}       \\ \hline
\multicolumn{1}{|c|}{Method}        & \multicolumn{1}{c|}{JCMSA} & \multicolumn{1}{c|}{The method in {\cite{wei2022robust}}} & 2D FFT method \\ \hline
\multicolumn{1}{|c|}{Run Time ($\mu$s)} & \multicolumn{1}{c|}{$2.401\times10^5$}   & \multicolumn{1}{c|}{$1.282\times10^3$}                    & 545.5          \\ \hline
\end{tabular}}
\end{center}
\end{table}

\section{Conclusion}\label{sc6}

For the ISAC-enabled mobile communication system operating in unlicensed spectrum bands,
conventional radar signal processing algorithms suffer from sidelobes deterioration and 
low anti-noise performance. 
To this end, the Joint CS and Machine Learning ISAC Signal Processing Algorithm (JCMSA) is proposed 
in this paper, which aims to mitigate sidelobes deterioration and 
enhance anti-noise performance. 
Specifically, by combining the RBG configuration information in 5G NR and CS techniques, which exploit the sparsity of the target to reconstruct the complete radar map, 
the enhanced range and velocity estimation algorithms are proposed.
Then, the FISTA algorithm is used to solve the convex optimization problem, 
while KCV is employed to select the regularization parameter to obtain zero sidelobes and better anti-noise performance than conventional 2D FFT algorithm.
Simulation results demonstrate that the proposed algorithm overcomes the deterioration of sidelobes. Meanwhile, the proposed algorithms in this paper have a maximum improvement of 54.66 \% and 84.36 \% in range estimation accuracy, and 41.54 \% and 97.09 \% in velocity estimation accuracy, compared with the improved 2D FFT algorithm in \cite{wei2022robust} and conventional 2D FFT algorithm, respectively.
With the emergence of various services, unlicensed spectrum bands have a large probability to be applied 
in ISAC-enabled mobile communication systems.
The work of this paper may provide beneficial references for the sensing algorithm design 
of the ISAC systems over unlicensed spectrum bands.

\bibliographystyle{IEEEtran}
\bibliography{Reference}

\ifCLASSOPTIONcaptionsoff
\newpage
\fi

\end{document}